%% file: LCNOMain.tex
\documentclass[reprint,amsmath,amssymb,aps,prb,superscriptaddress,floatfix]{revtex4-2}
\usepackage{natbib}
\usepackage{graphicx}
\usepackage{dcolumn}
\usepackage{bm}
\usepackage{float}
\usepackage{subfiles}
\usepackage{hyperref}
\usepackage[compact]{titlesec}
\usepackage{xcolor}
\begin{document}

\preprint{APS/123-QED}

\title{Spin re-orientation induced anisotropic magnetoresistance switching in LaCo$_{0.5}$Ni$_{0.5}$O$_{3-\delta}$ thin films}

\author{P. K. Sreejith}
\thanks{sreejithpksr@gmail.com}
\affiliation{Department of Physics, Low Temperature Physics Laboratory,  Indian Institute of Technology Madras, Chennai 600036, India}
\affiliation{Department of Physics, Nano Functional Materials Technology Center, Material Science Research Center, Indian Institute of Technology Madras, Chennai 600036, India}
\author{T. S. Suraj}
\affiliation{Department of Physics, National University of Singapore, 2 Science Drive 3, 117551, Singapore}
\author{Hari Babu Vasili}
\affiliation{School of Physics and Astronomy, W. H. Bragg Building, University of Leeds, Leeds LS2 9JT, UK}
\author{Suresh Sreya}
\affiliation{Department of Physics, Nano Functional Materials Technology Center, Material Science Research Center, Indian Institute of Technology Madras, Chennai 600036, India}

\author{Pierluigi Gargiani}
\affiliation{ALBA Synchrotron Light Source, E-08290 Cerdanyola del Vallès, Barcelona, Catalonia, Spain}
\author{K. Sethupathi}
\affiliation{Department of Physics, Low Temperature Physics Laboratory,  Indian Institute of Technology Madras, Chennai 600036, India}
\affiliation{Quantum Centre for Diamond and Emergent Materials (QuCenDiEM), Indian Institute of Technology Madras, Chennai 600036, India}
\author{Oscar Cespedes}
\affiliation{School of Physics and Astronomy, W. H. Bragg Building, University of Leeds, Leeds LS2 9JT, UK}
\author{V. Sankaranarayanan}
\affiliation{Department of Physics, Low Temperature Physics Laboratory, Indian Institute of Technology Madras, Chennai 600036, India}
\author{M. S. Ramachandra Rao}
\thanks{msrrao@iitm.ac.in}
\affiliation{Department of Physics, Nano Functional Materials Technology Center, Material Science Research Center, Indian Institute of Technology Madras, Chennai 600036, India}
\affiliation{Quantum Centre for Diamond and Emergent Materials (QuCenDiEM), Indian Institute of Technology Madras, Chennai 600036, India}
\date{\today}
\begin{abstract}
Realization of novel functionalities by tuning magnetic interactions in rare earth perovskite oxide thin films opens up exciting technological prospects. Strain-induced tuning of magnetic interactions in rare earth cobaltates and nickelates is of central importance due to their versatility in electronic transport properties. Here we report the spin re-orientation induced switching of anisotropic magnetoresistance (AMR) and its tunability with strain in epitaxial LaCo$_{0.5}$Ni$_{0.5}$O$_{3-\delta}$ thin films across the ferromagnetic transition. Moreover, with strain tuning, we observe a two-fold to four-fold symmetry crossover in AMR across the magnetic transition temperature. The magnetization measurements reveal a ferromagnetic transition around 50 K. At temperature below this transition, there is a subtle change in the magnetization dynamics, which reduces the  ferromagnetic long-range ordering in the system. X-ray absorption and X-ray magnetic circular dichroism spectroscopy measurements at the Co and Ni L edges reveal a Co spin state transition below 50 K, leading to the AMR switching and also the  presence of Ni$^{2+}$ and Co$^{4+}$ ions evidencing the charge transfer from Ni to Co ions. Our work demonstrated the tunability of magnetic interactions mediated electronic transport in cobaltate-nickelate thin films, which is relevant in understanding Ni-Co interactions in oxides for their technological applications such as in AMR sensors.
\end{abstract}

\maketitle
\section{Introduction}
\subfile{Introduction}
\section{Experimental Methods}
\input{Experimental_Methods}
\section{Results and Discussion}
\input{Results_and_discussion}
\section{Conclusion}
In summary, the present study elucidates the tunability of  anisotropic magnetoresistance with respect to temperature and strain in LCNO thin films arising from thermally induced Co spin state transitions. The observed reduction in magnetoconductance and inelastic phase coherence length below the magnetic transition temperature hints towards the onset of a glassy phase, similar to bulk LCNO, due to spin re-orientation. In addition, a direct correlation of the magnetic ordering to the magnetoresistance anisotropy can be seen across the ferromagnetic ordering temperature in the form of temperature-dependent sign reversal of AMR. The low temperature XAS and XMCD measurements performed across AMR switching temperatures shows Co$^{3+}$ spin state transition from HS/IS to LS, favouring antiferromagnetic super-exchange interaction. Further, we have observed a noticeable reduction in Co$^{4+}$ valence states for compressively strained films, which directly influenced the magnetic ordering in the system affecting the AMR behaviour. Compared to the parent LCO and LNO films, LCNO thin films hosts complex magnetic phase diagram involving multiple spin and valence states. The intricate magnetic ordering and spin re-orientation of LCNO triggered by thermally induced spin state transitions can be leveraged to achieve innovative functionalities in heterostructures, especially when combined with materials possessing high spin-orbit coupling.     

\begin{acknowledgments}
M.S.R. and K.S. would like to acknowledge the Science and Engineering Research Board (SERB) Grant No. EMR/2017/002328, Department of Science and Technology, Government of India (DST-GoI) funding which led to the establishment of the Nano Functional Materials Technology Centre (NFMTC) (Grants No. SR/NM/NAT/02-2005 and No. DST/NM/JIIT-01/2016(C)), DST FIST-Phase II funding for PPMS (SR/FST/PSII-038/2016 and IIT Madras for funding SVSM). We thank Dr. Manuel Valvidares, ALBA synchrotron light source, Spain for supporting the experiment and the beamtime.
\end{acknowledgments}

\section{\label{Appendix}Appendix}
\input{Appendix}

\bibliography{LCNOMain}

\end{document}

%% file: Introduction.tex
\begin{figure*}[!t]
\includegraphics[width=\linewidth]{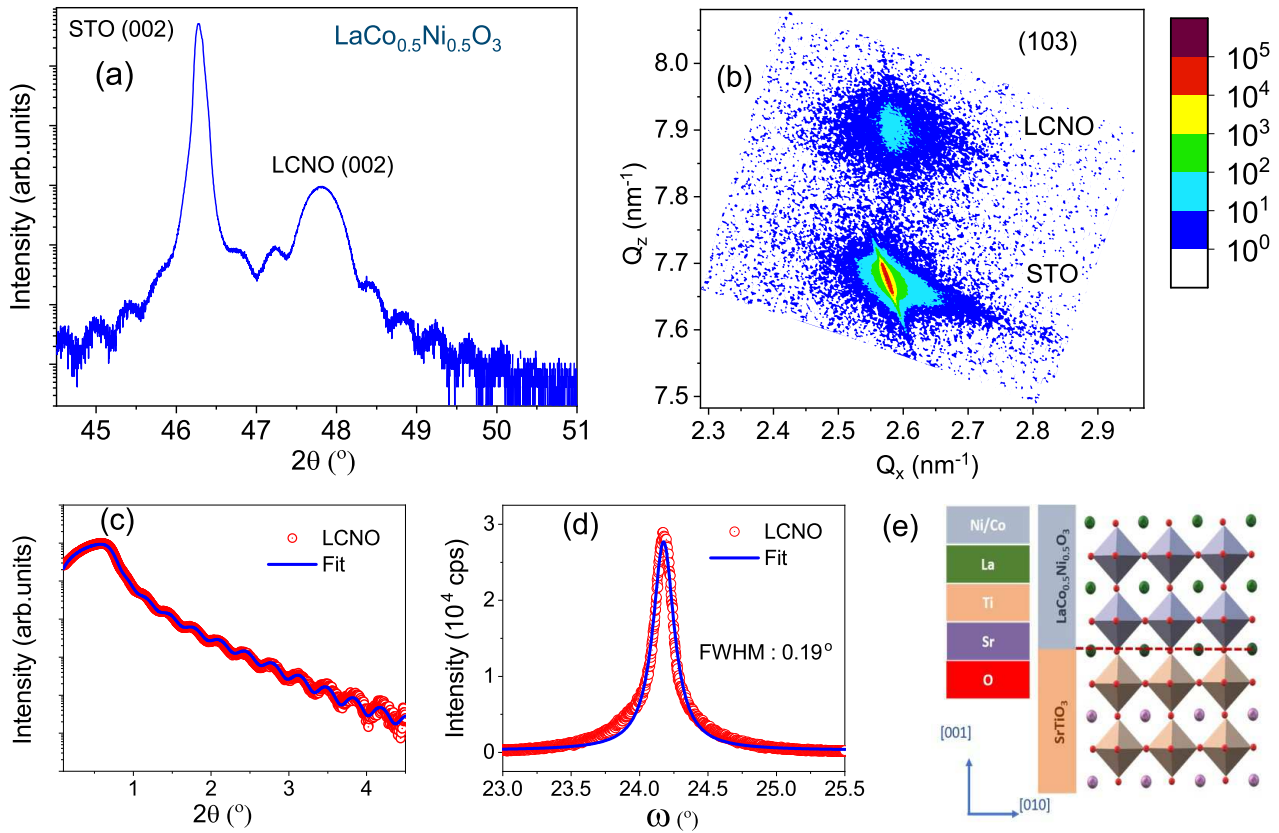}    
\centering
\caption{\label{Fig:xrd} \textbf{Structural characterization}. (a) Thin film XRD data of LCNO film on SrTiO$_3$ substrate with (002) peaks indexed. (b) RSM of the LCNO film along the (103) pseudo cubic plane. (c)Experimental XRR data (red open circles) for LCNO with blue line indicating the fit. From the fit, the film thickness and surface roughness were found to be ~25 nm and ~0.5 nm, respectively. (d) XRD $\omega$ rocking curve of LCNO (002) peak (red open circles)  with a Lorentzian fit (blue line). The FWHM is found to be $0.19^o$. (e)Schematic of LCNO stacking along the [001] direction. }
\end{figure*}
Transition metal oxides have generated large interest for their exciting magnetic and electrical properties such as superconductivity, colossal magnetoresistance (CMR), metal-insulator (M-I) transitions and large anisotropic magnetoresistance (AMR) \cite{Anderson2005}. AMR observed in ferromagnetic (FM) materials is usually associated with a change in the electron scattering rate depending upon the orientation of the magnetization with the applied current. \cite{Armitage2010,Ramirez1997,imada1998,Mcguire1975a,Egilmez2011a,Suraj2020,Vagadia2022}. AMR property has been utilized in spintronic applications and, more recently, in probing Fermi surface topology \cite{zhao2013research,sun2022anisotropic}

Transition metal oxides LaNiO$_3$ (LNO) and LaCoO$_3$ (LCO) are of particular interest for their charge, orbital and spin state correlations \cite{Sreedhar1992a,karolak2015correlation}. LNO is a highly correlated paramagnetic metal oxide, whereas LCO is a diamagnetic Mott insulator having temperature dependent multiple spin state transitions \cite{Rajeev1991,Sreedhar1992a,Xu}. In LCO, the non-magnetic insulating ground state is due to the fully occupied low-spin (LS) state of the trivalent cobalt ion. As the temperature  increases, the fully localized electrons in the $t_{2g}$ level undergo spin state transitions from a low-spin to an intermediate-spin (IS) state at about 90 K, and then to high-spin (HS) state beyond 500 K ~\cite{Heikes1964,Korotin1996,Raccah1967,Tokura1998,Naiman1965,Ren2011}. The temperature dependence of the Co$^{3+}$ spin states can be attributed to the comparable crystal field splitting energy (10 Dq) and the Hund’s exchange energy. 
These various ground states in LCO can be tuned by doping with other transition elements such as Ni, Mn and Fe \cite{Hammer2004,narasimhan1985structural}.

Solid solutions of rare earth cobaltate and nickelate perovskites are found to exhibit spin glass behaviour \cite{Viswanathan2009, Perez1998}. Spin glass behaviour arises due to competing FM and antiferromagnetic (AFM) interactions. 
Ferromagnetism arises from the double exchange interaction of Co$^{3+}$-O-Co$^{4+}$ whereas antiferromagnetism arises from the superexchange interactions arising from Co$^{3+}$-O-Co$^{3+}$ and Ni$^{3+}$-O-Ni$^{3+}$ \cite{Jonker1953MagneticCobalt}. Usually, large negative magnetoresistance is associated with spin glass systems because of the growth of FM domains in the presence of an external magnetic field which in turn promotes metallic conduction through the double exchange hopping mechanism \cite{Perez1998,motohashi2005}.
In thin-films, the influence of strain parameters further enhances the electrical transport and magnetic properties. Confinement effects have a profound impact on electronic transport, magnetoresistance (MR) and AMR \cite{Asaba2018}. The emergence of long-range ferromagnetism in thin film manganites is one example \cite{Millis1998a}. Tuning the electrical and magnetic properties of LCO by Ni doping in the Co site has been well studied in polycrystalline bulk form \cite{Hammer2004,Viswanathan2009}. However, studies on single crystals and epitaxial thin films are required to understand the role of magnetocrystalline anisotropy and strain effects in the electronic transport mechanism of Ni doped LCO system. 

 In this work, we discuss the microscopic origin of large positive magnetoconductance and anisotropic magnetoresistance switching in LaCo$_{0.5}$Ni$_{0.5}$O$_{3-\delta}$ (LCNO) thin films. Electrical transport measurements combined with X-ray absorption spectroscopy and magnetometry were used to elucidate the interplay of complex magnetic order and electron correlations. The manuscript is organized as the following sections: Section II deals with the experimental techniques used in this work. Section III consists of four parts. In part A, we discuss the structural and strain parameters of LCNO thin films. Part B deals with electrical transport properties, where we discuss the enhanced negative MR in epitaxially strained thin films. In part C, we discuss  angle-dependent magnetoresistance measurements. In section D, we discuss the magnetic properties, followed by the discussions on the electronic spin state transitions associated with the multivalent states of Co and Ni ions through XAS and XMCD measurements in section E.\\


%% file: Experimental_Methods.tex

\begin{figure*}[t]
\includegraphics[width=\linewidth]{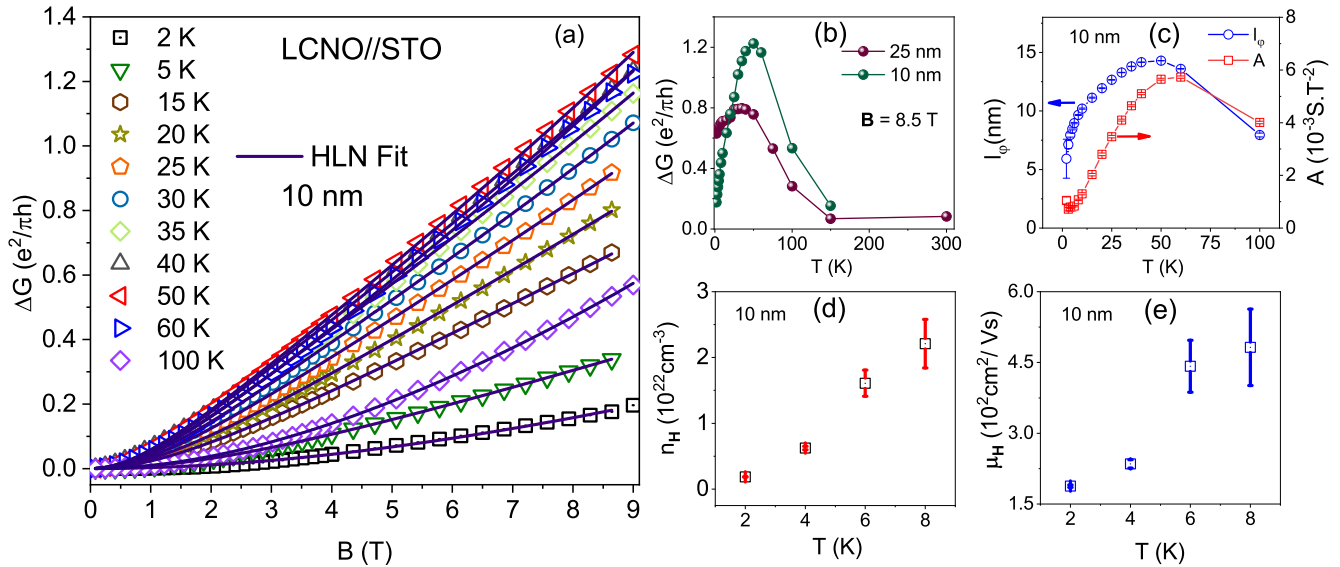}
\caption{\label{fig:delG} (a) \textbf{Magnetoconductance (a-c) and Hall effect measurements (d, e) }Experimental magnetoconductance ($\Delta G$) data  (open symbols) and its corresponding fit (solid lines) to the Hikami-Larkin-Nagaoka equation  with an additional electron-electron interaction term. (b) Temperature dependence of magnetoconductance for 25 nm and 10 nm films, respectively. (c) Inelastic scattering length ($l_\varphi$) and quadratic coefficient A, considering the $B^2$ dependence at high magnetic fields, both as a function of temperature. (d) Hall carrier concentration and (e) Hall carrier mobility as a function of temperature.}
\end{figure*}

Bulk polycrystalline LCNO target material for thin film deposition was synthesized using conventional citrate-based sol-gel technique \cite{nishio2018sol}. The final product thus obtained was cold pressed into a pellet and sintered at 900\textsuperscript{o}C for 24 h. Thin films of LCNO with various thicknesses were grown on TiO$_2$ terminated SrTiO$_3$ (001) (STO) substrates with the help of pulsed laser deposition (PLD) technique. A 248 nm KrF laser source was used with laser fluence of 2 Jcm$^{-2}$ and a pulse repetition rate of 3 Hz. The base pressure of the deposition chamber was set at 3 x $10^{-5}$ mbar and during deposition, a constant flow of high pure oxygen (99.999 \% ) O$_2$ with a partial pressure of 2 x 10$^{-1}$ mbar was maintained. The substrate temperature was kept at 700 $^o$C. Post deposition, 100 mbar O$_2$ partial pressure was maintained inside the deposition chamber while cooling down to room temperature to ensure proper oxygen stoichiometry.

The phase purity of thin film samples was confirmed using a high-resolution X-ray 
diffractometer (HRXRD) (Rigaku Smart Lab II) with a Cu K$_\alpha$ source ( $\lambda = 1.54$ \AA). $2\theta-\omega$ scans were done around the (200) peak of the STO substrate. 
From X-ray reflectivity (XRR) measurements, film thickness and roughness were determined. In order to study in-plane strain due to sample-substrate lattice mismatch, reciprocal space mapping (RSM) was carried out along the pseudocubic (103)$_{pc}$ crystal plane of the STO. Electrical transport measurements were carried out using a commercial Physical Property Measurement System (PPMS) (Quantum Design) from 300 K down to 2 K using the standard linear four-probe method. The contacts were aligned along the [010] direction of the crystal. Field-dependent measurements were carried out with varying magnetic fields up to 9 Tesla. Angle-dependent MR was carried out using the commercially available rotating holder attachment. Magnetization measurements were carried out in a SQUID based vibrating sample magnetometer (MPMS XL, Quantum Design). X-ray absorption spectroscopic measurements were carried out to investigate the magnetization and valence states of the transition elements. X-ray absorbtion spectrocsopy (XAS) and x-ray magnetic circular dichroism (XMCD) measurements were carried out at the BL-29 BOREAS beamline of the ALBA synchrotron light source \cite{barla2016design}. The XMCD spectra were obtained by taking the difference between the XAS of right ($\sigma^+$) and left ($\sigma^-$) circularly polarized light. The samples were measured in the in-plane geometry (grazing incidence from the film plane) under an applying magnetic field of 6 Tesla in the beam direction. The data were collected in the total electron yield (TEY) mode.

%% file: Results_and_discussion.tex
%
%

\subsection{Structural and morphological characterization}
Bulk LCNO has a rhombohedral (${R\bar3c}$) structure \cite{Viswanathan2009}. FIG.~\ref{Fig:xrd} (a) shows the room-temperature X-ray diffraction  $2\theta-\omega$ scan of the LCNO film grown on STO (001) substrate. The observation of Kiessig fringes in the $2\theta-\omega$ x-ray diffractogram confirms the good crystallinity of the films. To determine the film thickness, x-ray reflectometry (XRR) measurements were done and fitted with GenX software as shown in FIG.~\ref{Fig:xrd} (c) \cite{Bjorck2007}. The fit gives an average surface roughness (r$_a$) of $\sim$ 5\AA \: for as-grown films. The bulk pseudocubic (pc) lattice parameter for LCNO is found to be 3.789 \AA. On STO substrate, the out-of-plane lattice parameter of the LCNO thin film is found to be 3.801\AA. Further, from the RSM plots of (103)$_{pc}$ Bragg planes  (see FIG.~\ref{Fig:xrd} (b)), the in-plane and out-of-plane strain parameters relative to that of STO (a$_{pc}$= 3.90\AA) were found to be 0.25 \% and 2.8 \% tensile strained respectively for 25 nm thick film. A Rocking curve scan for the film shows a full width at half maximum (FWHM) of 0.19$ {\rm^o}$, again indicating good crystallinity of the film (shown in FIG.~\ref{Fig:xrd} (d)). The perovskite film is epitaxially oriented, grown along the [00l] direction of the STO substrate as shown in FIG.~\ref{Fig:xrd}).

\subsection{Electrical transport properties}
Temperature and magnetic field dependent electrical transport properties of LCNO films have been investigated down to 2 K. Temperature dependence of resistivity shows 3D Mott variable range hopping below 150 K. A detailed analysis is given in appendix-\ref{Appendix}. 

Magnetic field dependent resistivity measurements reveal large negative magnetoresistance in LCNO films. Though there can be multiple origins for negative magnetoresistance, in general, it can be attributed to the weak localization caused by the interference of the electron wavefunctions facilitated by  the time-reversal symmetry breaking due to the onset of ferromagnetic ordering below the transition temperature. Applying an external magnetic field will create a phase shift between the  electron wavefunctions and lift the localization. This results in a decrease in the resistivity of the system \cite{Wang1988}. Magnetoconductance ($\Delta$G) measurements were taken for LCNO thin films of thickness 10 nm and 25 nm, with field varying from -9 T to 9 T. The  magnetoconductance data were fitted with a simplified Hikami-Larkin-Nagaoka (HLN) equation within the spin-orbit coupling limit \cite{Hikami1980}. In the high field region, magnetoconductance is dominated by the elastic electron-electron scattering, and hence a quadratic $B^2$ field dependence is incorporated into the HLN equation, which thus can be written as
\begin{eqnarray}
\label{eqn:HLN}
\Delta G(B)&=&\frac{e^2}{2\pi^2\hbar}    
\left[\psi\left(\frac{1}{2}+\frac{B_\varphi}{B}\right)-ln\left(\frac{B_\varphi}{B}\right)\right]+A B^2\nonumber\\
\end{eqnarray}
where $\psi$ is digamma function, $B_\varphi$ is the effective field of inelastic scattering terms and is equal to $\left(\frac{\hbar}{4el_\varphi}\right)^2, l_\varphi$ is the phase coherence length and A is the electron-electron interaction coefficient.

\begin{figure*}[t]
    \centering
    \includegraphics[width=\linewidth,keepaspectratio]{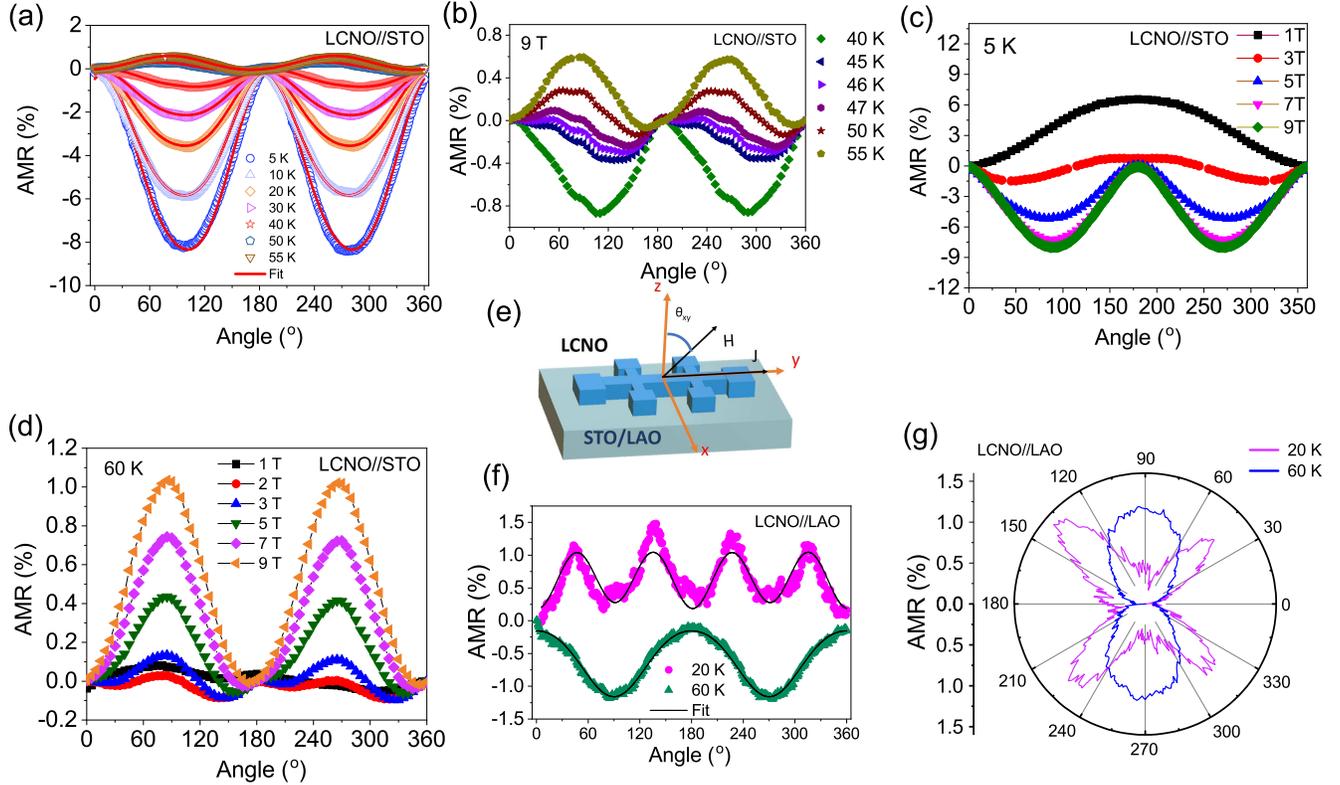}
\caption{\label{Fig:AMR} \textbf{AMR plots of LCNO films of 25 nm thickness grown on STO (a-d) and LAO (f, g) substrates.} (a) Temperature dependence of anisotropic magnetoresistance (AMR) of LCNO thin films for an applied field of 9 T, on STO substrate (open symbols), fitted with equation \ref{eqn:AMR3} (solid lines). (b) AMR data close to the transition temperature. (c) Field dependence of AMR below the magnetic transition. (d) Field dependence of AMR above the magnetic transition temperature. (e) Schematic drawing of the magnetic field rotation w.r.t. crystal plane and applied current (gamma configuration). (f) Temperature dependence of AMR in LAO substrate (with 9 T applied field). (g) Polar plots showing the two-fold to four-fold symmetry crossover in LCNO films grown on LAO.}
\end{figure*}

In FIG.~\ref{fig:delG} (a), the experimentally obtained conductivity of a 10 nm thin film has been fitted with equation~(\ref{eqn:HLN}). A magnetoconductance  maximum is obtained at the ordering temperature (see FIG. \ref{fig:delG} (b)).  FIG.~\ref{fig:delG} (c) gives the temperature dependence of inelastic phase coherence length, $l_\varphi$ and the electron-electron  interaction coefficient A. It is interesting to note that there is no monotonous increase or decrease in $l_\varphi$ and A. Instead, both parameters maximize near the ordering temperature and fall on either side. Correlation between the magnetoconductance maxima and $l_\varphi$ can be understood based on the ferromagnetic ordering of spins. The ferromagnetic alignment will reduce electron-electron scattering, which in turn increases $l_\varphi$. At low temperatures, some of the spins will re-orient themselves antiparallel and the long-range ordering is reduced, with the possible emergence of a glassy phase. It is known that in spin glass systems, the randomly fixed spins can dephase weak localization of electrons due to a reduction in magnetic scattering rate followed by suppression of the magnetic moment \cite{Wei1988,Peters1988,Schopfer2003}. Thus we see a reduction in $l_\varphi$ (see FIG.~\ref{fig:delG} (b)). A is added to the HLN equation to account for electron-electron elastic scattering and the classical cyclotron term. Based on the variations in the electron scattering cross-section due to ferromagnetic alignment followed by spin re-orientation, the term A follows a similar trend as $l_\varphi$ because of its direct correspondence to the electron scattering cross section variation.
Linear magnetoresistance data were taken for magnetic field sweeps from -9 T to +9 T. By taking the difference of the MR data in the two magnetic field sweep directions opposite to each other, we were able to deduce the magnetic field-dependent Hall resistance. By this procedure, in effect, we are removing the large linear magnetoresistance background. FIG.~\ref{fig:delG} (d) shows the Hall carrier concentration, n$_{H}$, which is extracted by fitting the slope of the linear region of Hall resistance,  and FIG.~\ref{fig:delG} (e) Hall carrier mobility, $\mu_H$ respectively with respect to temperature. The Hall carrier concentration is high, of the order of 10$^{22}$ cm$^{-3}$ which is similar to that of metals. However, the Hall mobility is very low, of the order of 10$^2$ cm$^2$V$^{-1}$s$^{-1}$, which can be associated to the localization of electrons.

\subsection{Anisotropic magnetoresistance switching}
\subsubsection*{\textbf{Origins of AMR in LCNO}}
AMR contains two components, crystalline and non-crystalline, where the former depends on the anisotropy of the electronic structure and spin-orbit coupling while the latter depends on the angle between the magnetization and current \cite{Rushforth2007}. The origins of the anisotropic magnetoresistance (AMR) can be directly correlated to the Fermi surface \cite{Mcguire1975a}. Applying a magnetic field will change the free electron trajectories near the anisotropic Fermi surface \cite{Kartsovnik1992}. For a uniformly magnetized system, the total free energy per unit volume as a function of the angle $\phi$ between the magnetisation and current direction can be given as \cite{Egilmez2011a}
\begin{eqnarray}
\label{eqn:AMR1}
E(\phi)&=& K_\text{shape}\sin^2{\phi}-\mu_0M_sH \cos{(\phi-\theta)}\nonumber\\
&&+K_\text{cry}\cos^2{(\phi-\alpha)}.
\end{eqnarray}
where \textbf{H} is the applied magnetic field, \textbf{M$_s$} is the saturation magnetization, $K_{shape}$ is the shape anisotropy constant related to the demagnetization factor, $K_{cry}$ is the magnetocrystalline anisotropic constant, $\alpha$ is the angle defining the crystallographic directions and $\theta$ is the angle between the applied magnetic field and current direction, which at high enough magnetic fields can be approximated to $\phi$. Minimizing the free energy and approximating $\phi$=$\theta$, one can obtain the anisotropic magnetoresistance as 
\begin{eqnarray}
\label{eqn:AMR2}
\rho_\text{AMR} = \rho_{\perp} + (\rho_{\parallel} - \rho_{\perp})\cos^2{\theta}.
\end{eqnarray}
 where $\rho_{\perp}$ and $\rho_{\parallel}$ are the resistivity at $\textbf{M}$ perpendicular and parallel to the applied current respectively. The AMR is therefore strongly dependent on cos$^2\theta$. This is generally valid for polycrystalline ferromagnetic materials. In the case of single crystalline and low dimensional epitaxial films with crystalline symmetry, magnetocrystalline effects need to be addressed \cite{Wu1253}. LCNO films grown along the [001] orientation have a cubic four-fold symmetry and an additional four-fold term representing the crystalline symmetry contributes towards AMR, thus modifying  the equation \ref{eqn:AMR2} \cite{Li2010a}
\begin{eqnarray}
\label{eqn:AMR3}
\rho_\text{AMR} = A_0 + A_2\cos[{2(\theta + \theta_0)}] + A_4\cos[{4(\theta + \theta_0)}]
\end{eqnarray}
where A$_2$ and A$_4$ are the respective amplitudes of two-fold and four-fold symmetry and $\theta_0$ accounts for the misalignment due to the sample rotator.
Angle-dependent magnetoresistance (ADMR) measurements were done on LCNO thin films in both gamma (B rotating along the yz plane where current I is along the y axis) and beta configurations (B rotation along the xz plane with current I in the y direction). FIG \ref{Fig:AMR} (a) shows the AMR, with an applied magnetic field of 9 T, at various temperatures taken by rotating the applied magnetic field direction with respect to the c-axis of the crystal. The $\%$ AMR is calculated as,
\begin{eqnarray}
\label{eqn:AMR3.1}
\frac{\rho(\theta)-\rho(\perp)}{\rho(\perp)}\times 100 \%
\end{eqnarray}
Here the magnetic field is applied along the [001] crystallographic direction. Both gamma and beta configurations show a similar trend in the AMR.

Temperature dependent ADMR studies have shown a large AMR in LCNO thin films (see FIG. \ref{Fig:AMR}). In the LaCo$_{1-x}$Ni$_x$O$_3$ system, the metal-insulator (M-I) transition occurs at x = 0.4 and hence LCNO, with x = 0.5, is close to this M-I transition and a large AMR is expected \cite{Hammer2004,Egilmez2011a}. In analogy to doped manganites, a disordered system like LCNO can be considered to have a conducting phase due to LNO and an insulating phase due to LCO. In thin films, due to film-substrate lattice mismatch, there will be an enhancement in the phase separation between these metallic and insulating phases. This, in turn, will enhance the anisotropic character of magnetoresistance  in the system. 
In the LCNO system, there is also a temperature dependent sign reversal of the AMR around 45 K from positive (above 50 K) to negative (below 40 K)-see FIG. \ref{Fig:AMR} (a) and (b). In FIG. \ref{Fig:AMR} (c) and (d) field-dependent variations of the AMR are shown at 5 K and 60 K (below and above the magnetic ordering temperature), respectively. The field-dependent evolution of AMR shows that only above 3 T magnetic field, the two-fold and four-fold components of the AMR sets in. At low fields, the demagnetization effects will destroy the effective AMR. Unlike some manganite systems, we could not observe field-dependent AMR switching \cite{Chen2009,Egilmez2011a}. 

\subsubsection*{\textbf{Spin polarized scattering and the sign of AMR}}
The temperature dependent sign reversal in AMR can be understood based on the change in the density of states of spin-up and spin-down conduction electrons above and below the magnetic transition. In ferromagnetic materials, due to spin polarization, the electron scattering cross section will be spin dependent. According to the microscopic theory, AMR can be defined as $\rho_\text{AMR} $= $\gamma(\beta - 1)$, where $\gamma$ is the spin-orbit coupling constant, which for most transition metals is about 0.01 and $\beta$ is the ratio of the spin-down to spin-up electron resistivity. Thus the sign of $\beta$ determines whether AMR is negative or positive \cite{Malozemoff1985,Campbell1970,Kokado2012a}. Hence in case of LCNO, switching of positive to negative AMR values as the temperature is lowered across the transition temperature can be understood as a shift in the density of states of spin-up and spin-down conduction electrons, which will affect their respective resistivities. A similar sign reversal in AMR is also observed in doped manganite systems \cite{Chen2009,Egilmez2011a}. 

\subsubsection*{\textbf{Low temperature spin-reorientation and AMR sign reversal}}
Below 50 K, there is a decrease in the net magnetization attributed to a possible spin re-orientation and the emergence of glassy phase well below 45 K as seen in bulk LCNO \cite{Viswanathan2009}. Above 45 K, ferromagnetism causes spin polarization and majority of electrons will be spin-up causing a reduction in the spin-up electron scattering leading to a positive AMR, which is in agreement with the magnetoconductance and magnetic data. Below 45 K, the long-range ferromagnetic ordering is lost due to spin re-orientation changing the spin polarization, and the AMR becomes negative. It should be noted that at low enough temperatures, some of the Co$^{3+}$ ions in LCNO have a spin state transition from HS/IS state to LS state as evidenced from the XMCD measurements on the Co L$_3$ edge. This will reduce the overall ferromagnetism in the system because in the LS state, the e$_g$ orbitals are vacant. These vacant e$_g$ orbitals will contribute towards the antiferromagnetic superexchange interaction and, in turn, will lead to the formation of a glassy phase in the system.

\subsubsection*{\textbf{Strain dependence of AMR}} 
The magnetocrystalline anisotropy and AMR are highly sensitive to strain. In order to understand the effects of strain on LCNO, we have also grown LCNO films ($\sim 25$ nm thick) on [001] oriented LAO substrates.  Compared to films grown on tensile STO substrates, films grown on LAO are compressively strained, which is evident from the XRD plot given in the appendix (see FIG. \ref{Fig:XRD2}). From FIG. \ref{Fig:AMR} (f) we can see a clear distinction from STO and LAO grown LCNO films with a crossover from two-fold to four-fold symmetry at 60 K and 20 K, respectively. FIG. \ref{Fig:AMR} (g) show the polar plot for the AMR shown in FIG. \ref{Fig:AMR}(f), where the two-fold to four-fold distinction is very evident. AMR studies on (GaMn)As epitaxial films have shown that the origin of fourfold symmetry is related to the long-range ferromagnetic ordering below the transition temperature, $T{\rm_C}$ \cite{Wu1253}. Further, in LCNO films grown on LAO, low temperature AMR at 20 K is positive while AMR at 60 K is negative. So we can infer that the spin dynamics are different for compressively grown LCNO films.
\begin{figure}[ht]
\includegraphics[width=\linewidth]{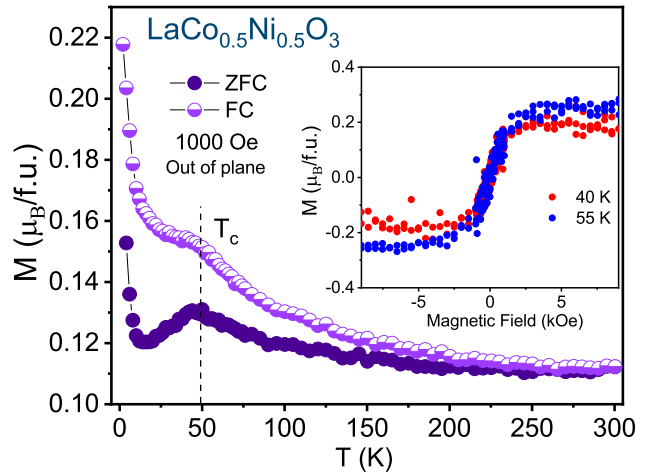}
\caption{\label{fig:MTMH} \textbf{Magnetisation measurements}. Zero field cooled (ZFC) and field cooled (FC) magnetization versus temperature data at 1000 Oe applied field. Inset shows the field dependent magnetization at 40 K and 55 K.}
\end{figure}

\begin{figure*}
\includegraphics[width=\linewidth]{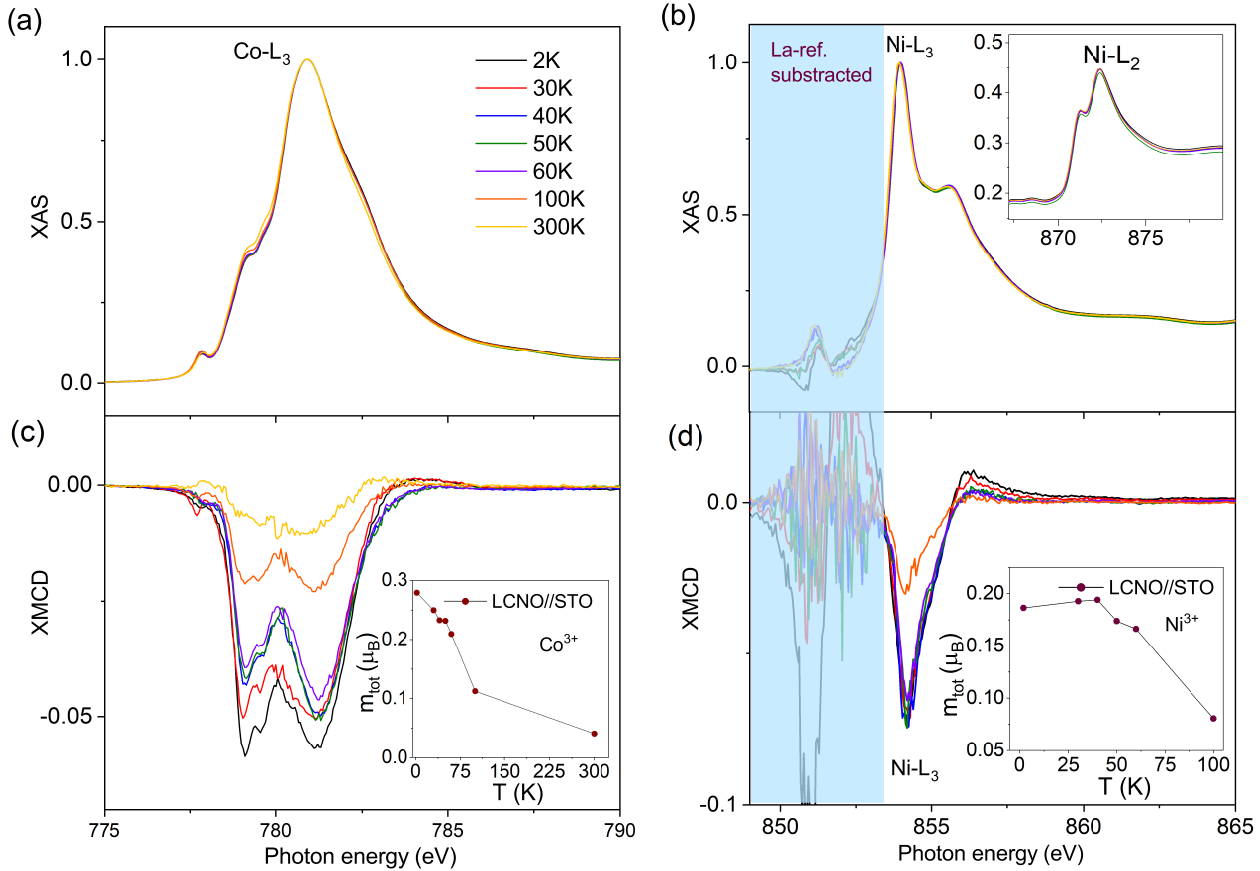}
\caption{\label{fig:XAS1} \textbf{X-ray absorption spectroscopy}. (a) XAS spectra of Co-L$_3$ edge at different temperatures. (b) XAS spectra of Ni-L$_3$ edge at different temperatures. The inset shows the XAS spectrum of Ni-L$_2$ edge. (c) XMCD spectra of Co-L$_3$ edge at different temperatures. Inset shows the temperature dependence of total magnetic moment derived from Co-L$_3$ edge XMCD using sum rules.(d) XMCD spectra of Ni-L3 edge at different temperatures with figure inset showing the temperature dependence of total magnetic moment derived from Ni-L3 edge XMCD using sum rules.}
\end{figure*}

\subsection {Magnetic properties}
Doping with trivalent transition metal ions in LCO is known to change the spin state of Co$^{3+}$ in the system. However, Ni$^{3+}$ in LNO remains at a low spin state throughout the temperature range, which is also the case for LCNO as evidenced from the XAS measurements given in the next section \cite{Kyomen2003,Kyomen2003b}. The presence of Ni ions in LCNO induces a change in Co$^{3+}$ spin state from low spin to intermediate spin or high spin state. The charge transfer between Ni and Co ions induces ferromagnetism in the system. Magnetization measurements, both temperature and field dependent, were done on LCNO films-see FIG. \ref{fig:MTMH}. The diamagnetic contribution from the STO substrates has been subtracted. FIG.~\ref{fig:MTMH} shows the temperature dependent magnetization when zero field cooled (ZFC) and field cooled (FC)), for a 25 nm thick LCNO film on STO substrate for an out-of-plane (OOP) 1000 Oe applied field. In line with reports of the properties of the bulk material, there is a ferromagnetic transition at around 50 K as evidenced by the temperature dependent magnetization measurements. However, as the temperature is lowered, the magnetization does not follow a Brillouin-like curve. Instead, the overall moment is reduced at low temperatures below 50 K, which could be due to possible spin re-orientation\cite{Viswanathan2009}. From the bifurcation of ZFC-FC curves, which indicates magnetic anisotropy, it is clear that the formation of the ferromagnetic cluster in the system starts way before the magnetic transition temperature in the system. This unusual magnetic order arises from Co to Ni charge transfer resulting in the formation of Co$^{4+}$ states and the Co$^{3+}$-O-Co$^{4+}$ double exchange interaction leading to ferromagnetism. To get further insights into spin dynamics across the ferromagnetic transition temperature, we have carried out temperature dependent X-ray absorption spectroscopy measurements.

\subsection{XAS and XMCD measurements}

\begin{figure}[ht]
\includegraphics[width=\linewidth]{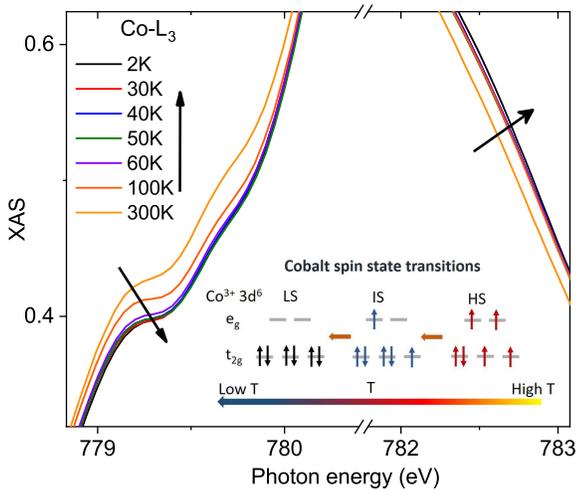}
\caption{\label{fig:Co-spin} \textbf{Spin state transition}. Enlarged view of Co-L$_3$ edge XAS spectra of FIG. \ref{fig:XAS1} showing the peak shift towards right with decrease in temperature. The black arrows are guide to the eyes. Inset shows the schematic of Co spin state transitions from HS (red) to IS (blue) and then to LS (black) with decrease in temperature.}
\end{figure}

XAS is a direct method to probe the electronic ground states and give better insights into the spin state transitions and valence states of cobalt and nickel ions in the LCNO thin film. The unusual temperature dependence of AMR phenomenon and their connection to the magnetic behaviour of the system can be addressed with temperature-dependent XMCD measurements. Using XAS and XMCD measurements at the Co-L$_{2,3}$ and Ni-L$_{2,3}$ absorption edges, we have probed the 3$d$ electronic ground states of Co and Ni ions in the system at various temperatures above and below the magnetic transition. The temperature dependence of XAS for Co-L$_3$  edge is shown in FIG. \ref{fig:XAS1} (a). The Co-L$_3$ main peak coincides with the Co$^{3+}$ valence ions, and the shoulder peaks to the left and right coincide with Co$^{4+}$ valency \cite{lin2010local,guillou2017valence}. Particularly, a double peak XAS feature at ~780 eV and intensified shoulder 
at 782.5 eV are not seen in the Co-L$_3$ XAS of pure LCO \cite{liao2019valence}. Hence, we confirm our samples are predominantly with Co$^{3+}$ state along with a possible Co$^{4+}$ state.
A small peak around 777 eV represents Co$^{2+}$ valence which could be due to the presence of oxygen vacancies mostly at the film surface \cite{chen2014complete}. Figure \ref{fig:Co-spin} shows the magnification of the XAS shoulder peaks, clearly showing a right shift of Co-L$_3$ peak as the temperature is reduced. With reference to the ground states of pure LS Co$^{3+}$ and pure HS Co$^{3+}$ compounds, the corresponding shifts can be inferred as a gradual transition from high spin (HS)/intermediate spin (IS) spin state at 300 K to low spin (LS) rich state at 2 K \cite{haverkort2006spin}.

The XMCD data in FIG. \ref{fig:XAS1} (c) shows a single peak at 300 K dominated by the HS/IS Co$^{3+}$ ions. As the temperature is reduced, the XMCD peak splits into a doublet with peaks at 779.1 and 781.2, corresponding to possible Co$^{3+}$ and Co$^{4+}$ ions respectively. The peak splitting is attributed to the charge transfer mediated increase in the Co valence. i.e. Co$^{4+}$, which favour ferromagnetic ordering and have transition corresponding to positive AMR at and above 50 K. We note that the peak intensity ratio Co$^{4+}$/Co$^{3+}$ increases and maximises at 50 K, coinciding with the ferromagnetic transition temperature obtained from the magnetic measurements using SQUID magnetometer. Importantly, the Co-L$_3$ XMCD (FIG. \ref{fig:XAS1}(c)) is found to be increasing with the decreasing temperature, resulting in a net increment in the Co magnetization. These qualitative changes in the XAS cannot complete the full picture of Co sub-lattice magnetization because the LCNO system contains multiple Co valent states either in HS or LS configurations. However, the spin ($m_s$) and orbital ($m_l$) magnetic moments derived from the XMCD data can provide further insight into the Co magnetization, and the discussion is in the following. 

We  have determined the m$_s$ and m$_l$ of Co (and Ni) ions by applying the sum rules to the integrated intensities of total XAS and XMCD spectra \cite{shen1995spatial}. Spin sum rule errors, associated to the core-valence exchange interactions, were properly accounted for the calculations \cite{piamonteze2009accuracy}. The most important part of the data is that the obtained $m_{tot} (= m_s + m_l)$ of Co ions from the XMCD data reveals a small but noticeable kink at 50 K as shown in the inset of FIG. \ref{fig:XAS1} (c), which is a fingerprint to the possible spin-reorientation transitions (due to the Co$^{4+}$) that already observed through bulk magnetization measured from the SQUID magnetometer, as shown in FIG. \ref{fig:MTMH}. All the magnetic moments reported here are for $\mu_B$/ion. The $m_l/m_s$ of Co is in the range of 0.45-0.53. These are slightly higher values than the pure Co$^{4+}$ (0.4-0.45) and lower than the pure Co$^{2+}$ (0.57) (\cite{guillou2017valence} and references there in). However, the $m_l/m_s$ varies between 0.2-0.5 for Co$^{3+}$ ions \cite{haverkort2006spin,merz2010x}. Our values are in good agreement within the possible range despite the $m_l/m_s$ of Co$^{3+}$ ions widely differ due to their complex nature of HS-LS switchover. We have furthermore investigated the Ni absorption edges; the data is discussed in the following. 

The XAS of Ni-L$_3$ pre-edge is overlapping with the La-M$_4$ post-edge. We have thus removed the La-M peak by subtracting the La-M XAS recorded on the LSMO reference sample, leaving out some subtracted traces in the Ni-L$_3$ pre-edge as shown in FIG. \ref{fig:XAS1} (b) for the temperature dependent Ni-L$_{2,3}$ XAS. However, the Ni-L$_2$ edge (FIG.\ref{fig:XAS1} (b) inset) has no overlap with the La M-edge, therefore, we clearly see the dominant peak at the higher energy side corresponds to Ni$^{3+}$ contribution whereas the Ni$^{2+}$ ion forming the shoulder peak left side to the main peak \cite{piamonteze2015interfacial}. Unlike Co-L$_3$ edge, the Ni-L$_3$ edge XAS is found to be independent of temperature and therefore no change in the spin state is expected. The Ni-L$_3$ XMCD (FIG. \ref{fig:XAS1}(d)) and further derived $m_{tot}$ value remain almost similar below 60 K (see inset of FIG. \ref{fig:XAS1} (b)). Above 50 K, the Ni moments follow a similar trend to that of Co ions. A total moment including Co and Ni ions is matching well with the magnetization data obtained from SQUID magnetometer.
\begin{figure}[t]
\includegraphics[width=\linewidth]{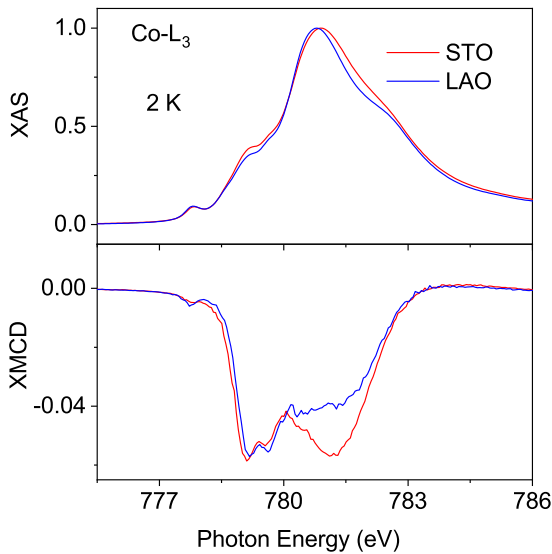}
\caption{\label{fig:XAS2} \textbf{Strain effects on LCNO: x-ray absorption spectrum}. XAS (top) and XMCD (bottom) spectrum of Co-L$_3$ edge of LCNO grown on STO (red) and LAO (blue) substrates}
\end{figure}
The effect of strain has a substantial influence on AMR in LCNO as shown in the previous section. To get a better understanding of the strain effects in AMR, temperature dependent XAS studies of LCNO films grown on STO and LAO substrates were compared. FIG. \ref{fig:XAS2} shows the XAS and XMCD spectra of Co-L$_3$ edge at 2 K of LCNO films grown on tensile strained STO and the compressively strained LAO substrates. The XAS spectrum of LAO-grown film is shifted towards the low energy side compared to its STO counterpart indicating a reduction in the Co$^{4+}$ valence state which is a clear indication of the variation of local spin magnetic moments with strain. \cite{freeland2008}
From the XMCD data in FIG. \ref{fig:XAS2} (a) we see a clear suppression of the peak at 781.5 eV for LAO-grown film indicating a reduction of Co$^{4+}$ spin states in the system.

X-ray absorption studies of Co and Ni edges confirm the role of cobalt spin-state transitions in tuning the magnetization in LCNO system while the valence and spin states of Ni ions remain same. Hence the AMR switching observed below T = 50 K can be attributed to the change in cobalt spin states, which contribute towards a net change in spin polarization densities by changing the magnetization dynamics across the magnetic transition temperature. The inset of FIG. \ref{fig:Co-spin} shows schematic representation of temperature dependent spin state transitions of Co$^{3+}$ ions. The high temperature region is dominated by HS state, followed by the IS state at intermediate temperatures and finally at low temperatures the concentration of LS state dominate.  The existence of Co$^{4+}$ and Ni$^{2+}$ valence states corroborates well with Ni to Co charge transfer of the type Ni$^{3+}$ + Co$^{3+} \rightarrow$ Ni$^{2+}$ + Co$^{4+}$ in LCNO films and the relative change in Co$^{4+}$/Co$^{3+}$ ratio, which is greatly modified by strain determines the extent of ferromagnetic ordering in the system.


%% file: Appendix.tex
%
\subsection*{Temperature dependence of electrical resistivity}

\begin{table}[b]
    \caption{Resistivity fit parameters for LCNO samples grown on SrTiO$_3$ substrates with various thicknesses}
    \centering
    \begin{ruledtabular}
    \begin{tabular}{ccccc}
    Film&\multicolumn{2}{c}{Mott VRH fit}&\multicolumn{2}{c}{Arrhenius fit} \\
    thickness\footnote{Films grown on SrTiO$_3$ substrate} (nm)&$\rho_o (\Omega.cm)$&$T_o (K)$&$\rho\textsuperscript{'}_o (\Omega.cm)$&$E_a(meV)$ \\ \hline\\
    $6$&$9.23\times10^{-9}$&$8.80\times10^6$&$2.02\times10^{-3}$&$44.9$\\
    $10$&$2.89\times10^{-5}$&$1.35\times10^5$&$1.76\times10^{-3}$&$18.2$\\
    $25$&$4.9\times10^{-4}$&$1.08\times10^4$&$3.48\times10^{-3}$&$12.0$\\ 
    \end{tabular}
    \end{ruledtabular}
    \label{tab:table1}
\end{table}
\begin{figure}[ht]
    \centering
    \includegraphics[width=\linewidth,keepaspectratio]{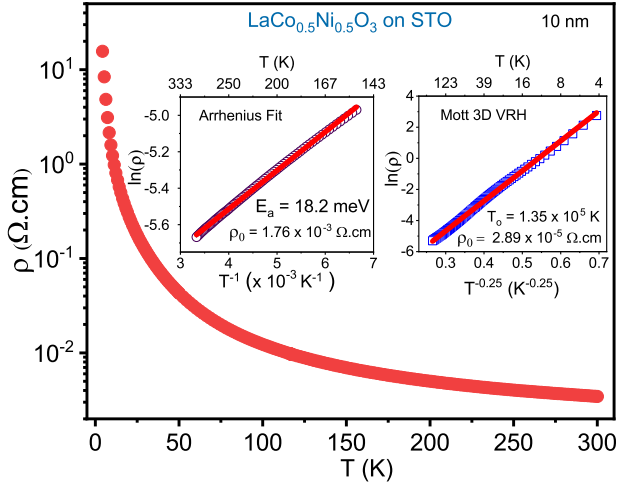}
\caption{\label{Fig:RT} \textbf{Electrical transport study}. Semi-log scale plot of resistivity versus temperature data for 10 nm LCNO film. Inset on the right is the 3D Mott VRH fit for low temperature resistivity data from 4 K to 150 K and inset on the left is the Arrhenius fit for the high temperature data from 150 K to 300 K.}
\end{figure}

LaNiO$_3$ is a paramagnetic metal throughout the temperature range and LaCoO$_3$ is a nonmagnetic insulator in its low temperature ground state. Therefore, from earlier studies, it is known that LaCo$_{0.5}$Ni$_{0.5}$O$_3$ falls near the metal-insulator transition (M-T) \cite{Hammer2004,Raychaudhuri1995}. Conduction mechanism in LaCo$_{1-x}$Ni$_x$O$_3$ is through O$_{2p}$ assisted d-d electronic transitions. At high temperatures, the resistivity follows the Arrhenius equation (see equation \ref{eqn:Arrh}, while for low enough temperatures, the conduction mechanism is governed by the Mott variable range hopping mechanism (equation (\ref{eqn:3DVRH})).

\begin{eqnarray}
\label{eqn:3DVRH}
\centering\rho(T)=\rho_o\exp
\left(
\frac{T_0}{T}
\right)^{1/n}
\end{eqnarray}
where $\rho_0$ is the exponential prefactor and $T_0$ is the Mott characteristic temperature defined as
\begin{eqnarray}
\centering T_o = \frac{18}{k_BN(E_F)\xi^3}
\end{eqnarray}
where $k_B$ is the Boltzmann constant, $N(E_F)$ is the density of states at the Fermi level and $\xi$ is the localization length.
Fig.\ref{Fig:RT} (a) shows the temperature dependence of resistivity for 10 nm LCNO sample. The resistivity plot shows a semiconducting behaviour with a steep increase in resistivity towards low temperatures. For temperatures from 4 K to 150 K,  resistivity data is fitted with equation (\ref{eqn:3DVRH}) (see Fig.\ref{Fig:RT} ~(b)). In the high-temperature region, i.e. from 150 K to 300 K, the resistivity data could be fitted by the phonon-assisted nearest neighbour hopping (NNH) interaction process, which is given by the Arrhenius equation as 
\begin{eqnarray}
\label{eqn:Arrh}
\centering\rho(T)=\rho\textsuperscript{'}_o\exp
\left(
\frac{E_a}{k_BT}
\right)
\end{eqnarray}
Further, we have also studied resistivity in films with varying thicknesses of 6 nm, 10 nm and 25 nm, and the corresponding fit parameters for equation \ref{eqn:3DVRH} and equation \ref{eqn:Arrh} are given in TABLE \ref{tab:table1}. From TABLE \ref{tab:table1}., we can see that as the thickness is reduced, activation energy is increased as inferred from the high-temperature Arrhenius fit.
\\
\vfill\eject
\subsection*{Thin film XRD of LCNO on LAO substrate}

\begin{figure}[h]
    \centering
    \includegraphics[width=\linewidth,keepaspectratio]{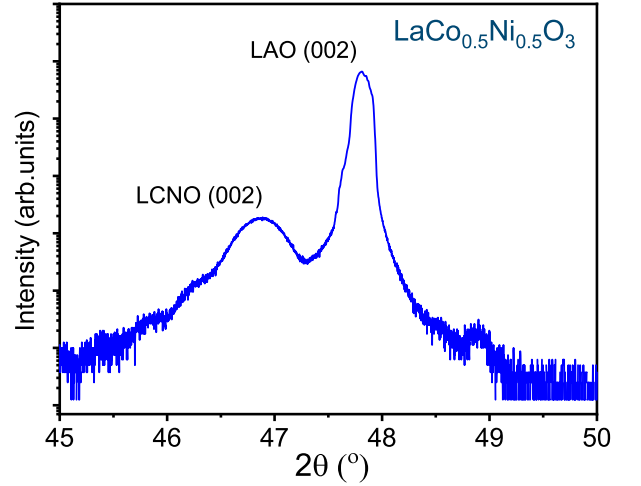}
\caption{\label{Fig:XRD2}Thin film XRD data of LCNO film on LaAlO$_3$ substrate with (002) peaks indexed.}
\end{figure}

%% file: LCNOMain.bbl
\begin{thebibliography}{57}%
\makeatletter
\providecommand \@ifxundefined [1]{%
 \@ifx{#1\undefined}
}%
\providecommand \@ifnum [1]{%
 \ifnum #1\expandafter \@firstoftwo
 \else \expandafter \@secondoftwo
 \fi
}%
\providecommand \@ifx [1]{%
 \ifx #1\expandafter \@firstoftwo
 \else \expandafter \@secondoftwo
 \fi
}%
\providecommand \natexlab [1]{#1}%
\providecommand \enquote  [1]{``#1''}%
\providecommand \bibnamefont  [1]{#1}%
\providecommand \bibfnamefont [1]{#1}%
\providecommand \citenamefont [1]{#1}%
\providecommand \href@noop [0]{\@secondoftwo}%
\providecommand \href [0]{\begingroup \@sanitize@url \@href}%
\providecommand \@href[1]{\@@startlink{#1}\@@href}%
\providecommand \@@href[1]{\endgroup#1\@@endlink}%
\providecommand \@sanitize@url [0]{\catcode `\\12\catcode `\$12\catcode
  `\&12\catcode `\#12\catcode `\^12\catcode `\_12\catcode `\%12\relax}%
\providecommand \@@startlink[1]{}%
\providecommand \@@endlink[0]{}%
\providecommand \url  [0]{\begingroup\@sanitize@url \@url }%
\providecommand \@url [1]{\endgroup\@href {#1}{\urlprefix }}%
\providecommand \urlprefix  [0]{URL }%
\providecommand \Eprint [0]{\href }%
\providecommand \doibase [0]{https://doi.org/}%
\providecommand \selectlanguage [0]{\@gobble}%
\providecommand \bibinfo  [0]{\@secondoftwo}%
\providecommand \bibfield  [0]{\@secondoftwo}%
\providecommand \translation [1]{[#1]}%
\providecommand \BibitemOpen [0]{}%
\providecommand \bibitemStop [0]{}%
\providecommand \bibitemNoStop [0]{.\EOS\space}%
\providecommand \EOS [0]{\spacefactor3000\relax}%
\providecommand \BibitemShut  [1]{\csname bibitem#1\endcsname}%
\let\auto@bib@innerbib\@empty
\bibitem [{\citenamefont {Anderson}\ and\ \citenamefont
  {Edwardst}(2005)}]{Anderson2005}%
  \BibitemOpen
  \bibfield  {author} {\bibinfo {author} {\bibfnamefont {P.~W.}\ \bibnamefont
  {Anderson}}\ and\ \bibinfo {author} {\bibfnamefont {S.~F.}\ \bibnamefont
  {Edwardst}},\ }\bibfield  {booktitle} {\emph {\bibinfo {booktitle} {A Career
  in Theoretical Physics, (2nd Edition)}},\ }\href
  {https://doi.org/10.1142/9789812567154\_0039} {\bibfield  {journal} {\bibinfo
   {journal} {World Scientific Series in 20th Century Physics}\ }\textbf
  {\bibinfo {volume} {965}},\ \bibinfo {pages} {510} (\bibinfo {year}
  {2005})}\BibitemShut {NoStop}%
\bibitem [{\citenamefont {Armitage}\ \emph {et~al.}(2010)\citenamefont
  {Armitage}, \citenamefont {Fournier},\ and\ \citenamefont
  {Greene}}]{Armitage2010}%
  \BibitemOpen
  \bibfield  {author} {\bibinfo {author} {\bibfnamefont {N.~P.}\ \bibnamefont
  {Armitage}}, \bibinfo {author} {\bibfnamefont {P.}~\bibnamefont {Fournier}},\
  and\ \bibinfo {author} {\bibfnamefont {R.~L.}\ \bibnamefont {Greene}},\
  }\href {https://doi.org/10.1103/RevModPhys.82.2421} {\bibfield  {journal}
  {\bibinfo  {journal} {Rev. Mod. Phys.}\ }\textbf {\bibinfo {volume} {82}},\
  \bibinfo {pages} {2421} (\bibinfo {year} {2010})}\BibitemShut {NoStop}%
\bibitem [{\citenamefont {Ramirez}(1997)}]{Ramirez1997}%
  \BibitemOpen
  \bibfield  {author} {\bibinfo {author} {\bibfnamefont {A.~P.}\ \bibnamefont
  {Ramirez}},\ }\href {https://doi.org/10.1088/0953-8984/9/39/005} {\bibinfo
  {title} {Colossal magnetoresistance}} (\bibinfo {year} {1997})\BibitemShut
  {NoStop}%
\bibitem [{\citenamefont {Imada}\ \emph {et~al.}(1998)\citenamefont {Imada},
  \citenamefont {Fujimori},\ and\ \citenamefont {Tokura}}]{imada1998}%
  \BibitemOpen
  \bibfield  {author} {\bibinfo {author} {\bibfnamefont {M.}~\bibnamefont
  {Imada}}, \bibinfo {author} {\bibfnamefont {A.}~\bibnamefont {Fujimori}},\
  and\ \bibinfo {author} {\bibfnamefont {Y.}~\bibnamefont {Tokura}},\ }\href
  {https://doi.org/10.1103/revmodphys.70.1039} {\bibfield  {journal} {\bibinfo
  {journal} {Rev. Mod. Phys.}\ }\textbf {\bibinfo {volume} {70}},\ \bibinfo
  {pages} {1039} (\bibinfo {year} {1998})}\BibitemShut {NoStop}%
\bibitem [{\citenamefont {Mcguire}\ and\ \citenamefont
  {Potter}(1975)}]{Mcguire1975a}%
  \BibitemOpen
  \bibfield  {author} {\bibinfo {author} {\bibfnamefont {T.~R.}\ \bibnamefont
  {Mcguire}}\ and\ \bibinfo {author} {\bibfnamefont {R.~I.}\ \bibnamefont
  {Potter}},\ }\href {https://doi.org/10.1109/TMAG.1975.1058782} {\bibfield
  {journal} {\bibinfo  {journal} {IEEE Trans. Magn.}\ }\textbf {\bibinfo
  {volume} {11}},\ \bibinfo {pages} {1018} (\bibinfo {year}
  {1975})}\BibitemShut {NoStop}%
\bibitem [{\citenamefont {Egilmez}\ \emph {et~al.}(2011)\citenamefont
  {Egilmez}, \citenamefont {Chow},\ and\ \citenamefont {Jung}}]{Egilmez2011a}%
  \BibitemOpen
  \bibfield  {author} {\bibinfo {author} {\bibfnamefont {M.}~\bibnamefont
  {Egilmez}}, \bibinfo {author} {\bibfnamefont {K.~H.}\ \bibnamefont {Chow}},\
  and\ \bibinfo {author} {\bibfnamefont {J.~A.}\ \bibnamefont {Jung}},\ }\href
  {https://doi.org/10.1142/S0217984911026176} {\bibfield  {journal} {\bibinfo
  {journal} {Mod. Phys. Lett. B}\ }\textbf {\bibinfo {volume} {25}},\ \bibinfo
  {pages} {697} (\bibinfo {year} {2011})}\BibitemShut {NoStop}%
\bibitem [{\citenamefont {Suraj}\ \emph {et~al.}(2020)\citenamefont {Suraj},
  \citenamefont {Omar}, \citenamefont {Jani}, \citenamefont {Juvaid},
  \citenamefont {Hooda}, \citenamefont {Chaudhuri}, \citenamefont {Rusydi},
  \citenamefont {Sethupathi}, \citenamefont {Venkatesan}, \citenamefont
  {Ariando},\ and\ \citenamefont {Rao}}]{Suraj2020}%
  \BibitemOpen
  \bibfield  {author} {\bibinfo {author} {\bibfnamefont {T.~S.}\ \bibnamefont
  {Suraj}}, \bibinfo {author} {\bibfnamefont {G.~J.}\ \bibnamefont {Omar}},
  \bibinfo {author} {\bibfnamefont {H.}~\bibnamefont {Jani}}, \bibinfo {author}
  {\bibfnamefont {M.~M.}\ \bibnamefont {Juvaid}}, \bibinfo {author}
  {\bibfnamefont {S.}~\bibnamefont {Hooda}}, \bibinfo {author} {\bibfnamefont
  {A.}~\bibnamefont {Chaudhuri}}, \bibinfo {author} {\bibfnamefont
  {A.}~\bibnamefont {Rusydi}}, \bibinfo {author} {\bibfnamefont
  {K.}~\bibnamefont {Sethupathi}}, \bibinfo {author} {\bibfnamefont
  {T.}~\bibnamefont {Venkatesan}}, \bibinfo {author} {\bibfnamefont
  {A.}~\bibnamefont {Ariando}},\ and\ \bibinfo {author} {\bibfnamefont
  {M.~S.~R.}\ \bibnamefont {Rao}},\ }\href
  {https://doi.org/10.1103/PhysRevB.102.125145} {\bibfield  {journal} {\bibinfo
   {journal} {Phys. Rev. B}\ }\textbf {\bibinfo {volume} {102}},\ \bibinfo
  {pages} {125145} (\bibinfo {year} {2020})}\BibitemShut {NoStop}%
\bibitem [{\citenamefont {Vagadia}\ \emph {et~al.}(2022)\citenamefont
  {Vagadia}, \citenamefont {Sardar}, \citenamefont {Tank}, \citenamefont {Das},
  \citenamefont {Gunn}, \citenamefont {Pandey}, \citenamefont {H\"ubner},
  \citenamefont {Rodolakis}, \citenamefont {Fabbris}, \citenamefont {Choi},
  \citenamefont {Haskel}, \citenamefont {Frano},\ and\ \citenamefont
  {Rana}}]{Vagadia2022}%
  \BibitemOpen
  \bibfield  {author} {\bibinfo {author} {\bibfnamefont {M.}~\bibnamefont
  {Vagadia}}, \bibinfo {author} {\bibfnamefont {S.}~\bibnamefont {Sardar}},
  \bibinfo {author} {\bibfnamefont {T.}~\bibnamefont {Tank}}, \bibinfo {author}
  {\bibfnamefont {S.}~\bibnamefont {Das}}, \bibinfo {author} {\bibfnamefont
  {B.}~\bibnamefont {Gunn}}, \bibinfo {author} {\bibfnamefont {P.}~\bibnamefont
  {Pandey}}, \bibinfo {author} {\bibfnamefont {R.}~\bibnamefont {H\"ubner}},
  \bibinfo {author} {\bibfnamefont {F.}~\bibnamefont {Rodolakis}}, \bibinfo
  {author} {\bibfnamefont {G.}~\bibnamefont {Fabbris}}, \bibinfo {author}
  {\bibfnamefont {Y.}~\bibnamefont {Choi}}, \bibinfo {author} {\bibfnamefont
  {D.}~\bibnamefont {Haskel}}, \bibinfo {author} {\bibfnamefont
  {A.}~\bibnamefont {Frano}},\ and\ \bibinfo {author} {\bibfnamefont {D.~S.}\
  \bibnamefont {Rana}},\ }\href {https://doi.org/10.1103/PhysRevB.105.L020402}
  {\bibfield  {journal} {\bibinfo  {journal} {Phys. Rev. B}\ }\textbf {\bibinfo
  {volume} {105}},\ \bibinfo {pages} {L020402} (\bibinfo {year}
  {2022})}\BibitemShut {NoStop}%
\bibitem [{\citenamefont {Zhao}\ \emph {et~al.}(2013)\citenamefont {Zhao},
  \citenamefont {Ding}, \citenamefont {HuangFu}, \citenamefont {Zhang},\ and\
  \citenamefont {Yu}}]{zhao2013research}%
  \BibitemOpen
  \bibfield  {author} {\bibinfo {author} {\bibfnamefont {C.-J.}\ \bibnamefont
  {Zhao}}, \bibinfo {author} {\bibfnamefont {L.}~\bibnamefont {Ding}}, \bibinfo
  {author} {\bibfnamefont {J.-S.}\ \bibnamefont {HuangFu}}, \bibinfo {author}
  {\bibfnamefont {J.-Y.}\ \bibnamefont {Zhang}},\ and\ \bibinfo {author}
  {\bibfnamefont {G.-H.}\ \bibnamefont {Yu}},\ }\href@noop {} {\bibfield
  {journal} {\bibinfo  {journal} {Rare Met.}\ }\textbf {\bibinfo {volume}
  {32}},\ \bibinfo {pages} {213} (\bibinfo {year} {2013})}\BibitemShut
  {NoStop}%
\bibitem [{\citenamefont {Sun}\ \emph {et~al.}(2022)\citenamefont {Sun},
  \citenamefont {Tang}, \citenamefont {Shen}, \citenamefont {Sun},
  \citenamefont {Zhao}, \citenamefont {Han}, \citenamefont {Kan}, \citenamefont
  {Cong}, \citenamefont {Zhang}, \citenamefont {Han}, \citenamefont {Qian},
  \citenamefont {Jiang}, \citenamefont {Wang}, \citenamefont {Zheng},\ and\
  \citenamefont {Fang}}]{sun2022anisotropic}%
  \BibitemOpen
  \bibfield  {author} {\bibinfo {author} {\bibfnamefont {X.}~\bibnamefont
  {Sun}}, \bibinfo {author} {\bibfnamefont {F.}~\bibnamefont {Tang}}, \bibinfo
  {author} {\bibfnamefont {X.}~\bibnamefont {Shen}}, \bibinfo {author}
  {\bibfnamefont {W.}~\bibnamefont {Sun}}, \bibinfo {author} {\bibfnamefont
  {W.}~\bibnamefont {Zhao}}, \bibinfo {author} {\bibfnamefont {Y.}~\bibnamefont
  {Han}}, \bibinfo {author} {\bibfnamefont {X.}~\bibnamefont {Kan}}, \bibinfo
  {author} {\bibfnamefont {S.}~\bibnamefont {Cong}}, \bibinfo {author}
  {\bibfnamefont {L.}~\bibnamefont {Zhang}}, \bibinfo {author} {\bibfnamefont
  {Z.}~\bibnamefont {Han}}, \bibinfo {author} {\bibfnamefont {B.}~\bibnamefont
  {Qian}}, \bibinfo {author} {\bibfnamefont {X.}~\bibnamefont {Jiang}},
  \bibinfo {author} {\bibfnamefont {S.}~\bibnamefont {Wang}}, \bibinfo {author}
  {\bibfnamefont {R.}~\bibnamefont {Zheng}},\ and\ \bibinfo {author}
  {\bibfnamefont {Y.}~\bibnamefont {Fang}},\ }\href
  {https://doi.org/10.1103/PhysRevB.105.195114} {\bibfield  {journal} {\bibinfo
   {journal} {Phys. Rev. B}\ }\textbf {\bibinfo {volume} {105}},\ \bibinfo
  {pages} {195114} (\bibinfo {year} {2022})}\BibitemShut {NoStop}%
\bibitem [{\citenamefont {Sreedhar}\ \emph {et~al.}(1992)\citenamefont
  {Sreedhar}, \citenamefont {Honig}, \citenamefont {Darwin}, \citenamefont
  {McElfresh}, \citenamefont {Shand}, \citenamefont {Xu}, \citenamefont
  {Crooker},\ and\ \citenamefont {Spalek}}]{Sreedhar1992a}%
  \BibitemOpen
  \bibfield  {author} {\bibinfo {author} {\bibfnamefont {K.}~\bibnamefont
  {Sreedhar}}, \bibinfo {author} {\bibfnamefont {J.~M.}\ \bibnamefont {Honig}},
  \bibinfo {author} {\bibfnamefont {M.}~\bibnamefont {Darwin}}, \bibinfo
  {author} {\bibfnamefont {M.}~\bibnamefont {McElfresh}}, \bibinfo {author}
  {\bibfnamefont {P.~M.}\ \bibnamefont {Shand}}, \bibinfo {author}
  {\bibfnamefont {J.}~\bibnamefont {Xu}}, \bibinfo {author} {\bibfnamefont
  {B.~C.}\ \bibnamefont {Crooker}},\ and\ \bibinfo {author} {\bibfnamefont
  {J.}~\bibnamefont {Spalek}},\ }\href
  {https://doi.org/10.1103/PhysRevB.46.6382} {\bibfield  {journal} {\bibinfo
  {journal} {Phys. Rev. B}\ }\textbf {\bibinfo {volume} {46}},\ \bibinfo
  {pages} {6382} (\bibinfo {year} {1992})}\BibitemShut {NoStop}%
\bibitem [{\citenamefont {Karolak}\ \emph {et~al.}(2015)\citenamefont
  {Karolak}, \citenamefont {Izquierdo}, \citenamefont {Molodtsov},\ and\
  \citenamefont {Lichtenstein}}]{karolak2015correlation}%
  \BibitemOpen
  \bibfield  {author} {\bibinfo {author} {\bibfnamefont {M.}~\bibnamefont
  {Karolak}}, \bibinfo {author} {\bibfnamefont {M.}~\bibnamefont {Izquierdo}},
  \bibinfo {author} {\bibfnamefont {S.~L.}\ \bibnamefont {Molodtsov}},\ and\
  \bibinfo {author} {\bibfnamefont {A.~I.}\ \bibnamefont {Lichtenstein}},\
  }\href {https://doi.org/10.1103/PhysRevLett.115.046401} {\bibfield  {journal}
  {\bibinfo  {journal} {Phys. Rev. Lett.}\ }\textbf {\bibinfo {volume} {115}},\
  \bibinfo {pages} {046401} (\bibinfo {year} {2015})}\BibitemShut {NoStop}%
\bibitem [{\citenamefont {Rajeev}\ \emph {et~al.}(1991)\citenamefont {Rajeev},
  \citenamefont {Shivashankar},\ and\ \citenamefont
  {Raychaudhuri}}]{Rajeev1991}%
  \BibitemOpen
  \bibfield  {author} {\bibinfo {author} {\bibfnamefont {K.~P.}\ \bibnamefont
  {Rajeev}}, \bibinfo {author} {\bibfnamefont {G.~V.}\ \bibnamefont
  {Shivashankar}},\ and\ \bibinfo {author} {\bibfnamefont {A.~K.}\ \bibnamefont
  {Raychaudhuri}},\ }\href {https://doi.org/10.1016/0038-1098(91)90915-I}
  {\bibfield  {journal} {\bibinfo  {journal} {Solid State Commun.}\ }\textbf
  {\bibinfo {volume} {79}},\ \bibinfo {pages} {591} (\bibinfo {year}
  {1991})}\BibitemShut {NoStop}%
\bibitem [{\citenamefont {Xu}\ \emph {et~al.}(1993)\citenamefont {Xu},
  \citenamefont {Peng}, \citenamefont {Li}, \citenamefont {Ju},\ and\
  \citenamefont {Greene}}]{Xu}%
  \BibitemOpen
  \bibfield  {author} {\bibinfo {author} {\bibfnamefont {X.~Q.}\ \bibnamefont
  {Xu}}, \bibinfo {author} {\bibfnamefont {J.~L.}\ \bibnamefont {Peng}},
  \bibinfo {author} {\bibfnamefont {Z.~Y.}\ \bibnamefont {Li}}, \bibinfo
  {author} {\bibfnamefont {H.~L.}\ \bibnamefont {Ju}},\ and\ \bibinfo {author}
  {\bibfnamefont {R.~L.}\ \bibnamefont {Greene}},\ }\href
  {https://doi.org/10.1103/PhysRevB.48.1112} {\bibfield  {journal} {\bibinfo
  {journal} {Phys. Rev. B}\ }\textbf {\bibinfo {volume} {48}},\ \bibinfo
  {pages} {1112} (\bibinfo {year} {1993})}\BibitemShut {NoStop}%
\bibitem [{\citenamefont {Heikes}\ \emph {et~al.}(1964)\citenamefont {Heikes},
  \citenamefont {Miller},\ and\ \citenamefont {Mazelsky}}]{Heikes1964}%
  \BibitemOpen
  \bibfield  {author} {\bibinfo {author} {\bibfnamefont {R.~R.}\ \bibnamefont
  {Heikes}}, \bibinfo {author} {\bibfnamefont {R.~C.}\ \bibnamefont {Miller}},\
  and\ \bibinfo {author} {\bibfnamefont {R.}~\bibnamefont {Mazelsky}},\ }\href
  {https://doi.org/10.1016/0031-8914(64)90182-X} {\bibfield  {journal}
  {\bibinfo  {journal} {Physica}\ }\textbf {\bibinfo {volume} {30}},\ \bibinfo
  {pages} {1600} (\bibinfo {year} {1964})}\BibitemShut {NoStop}%
\bibitem [{\citenamefont {Korotin}\ \emph {et~al.}(1996)\citenamefont
  {Korotin}, \citenamefont {Ezhov}, \citenamefont {Solovyev}, \citenamefont
  {Anisimov}, \citenamefont {Khomskii},\ and\ \citenamefont
  {Sawatzky}}]{Korotin1996}%
  \BibitemOpen
  \bibfield  {author} {\bibinfo {author} {\bibfnamefont {M.~A.}\ \bibnamefont
  {Korotin}}, \bibinfo {author} {\bibfnamefont {S.~Y.}\ \bibnamefont {Ezhov}},
  \bibinfo {author} {\bibfnamefont {I.~V.}\ \bibnamefont {Solovyev}}, \bibinfo
  {author} {\bibfnamefont {V.~I.}\ \bibnamefont {Anisimov}}, \bibinfo {author}
  {\bibfnamefont {D.~I.}\ \bibnamefont {Khomskii}},\ and\ \bibinfo {author}
  {\bibfnamefont {G.~A.}\ \bibnamefont {Sawatzky}},\ }\href
  {https://doi.org/10.1103/PhysRevB.54.5309} {\bibfield  {journal} {\bibinfo
  {journal} {Phys. Rev. B}\ }\textbf {\bibinfo {volume} {54}},\ \bibinfo
  {pages} {5309} (\bibinfo {year} {1996})}\BibitemShut {NoStop}%
\bibitem [{\citenamefont {Raccah}\ and\ \citenamefont
  {Goodenough}(1967)}]{Raccah1967}%
  \BibitemOpen
  \bibfield  {author} {\bibinfo {author} {\bibfnamefont {P.~M.}\ \bibnamefont
  {Raccah}}\ and\ \bibinfo {author} {\bibfnamefont {J.~B.}\ \bibnamefont
  {Goodenough}},\ }\href {https://doi.org/10.1103/PhysRev.155.932} {\bibfield
  {journal} {\bibinfo  {journal} {Phys. Rev.}\ }\textbf {\bibinfo {volume}
  {103}},\ \bibinfo {pages} {626} (\bibinfo {year} {1967})}\BibitemShut
  {NoStop}%
\bibitem [{\citenamefont {Tokura}\ \emph {et~al.}(1998)\citenamefont {Tokura},
  \citenamefont {Okimoto}, \citenamefont {Yamaguchi}, \citenamefont
  {Taniguchi}, \citenamefont {Kimura},\ and\ \citenamefont
  {Takagi}}]{Tokura1998}%
  \BibitemOpen
  \bibfield  {author} {\bibinfo {author} {\bibfnamefont {Y.}~\bibnamefont
  {Tokura}}, \bibinfo {author} {\bibfnamefont {Y.}~\bibnamefont {Okimoto}},
  \bibinfo {author} {\bibfnamefont {S.}~\bibnamefont {Yamaguchi}}, \bibinfo
  {author} {\bibfnamefont {H.}~\bibnamefont {Taniguchi}}, \bibinfo {author}
  {\bibfnamefont {T.}~\bibnamefont {Kimura}},\ and\ \bibinfo {author}
  {\bibfnamefont {H.}~\bibnamefont {Takagi}},\ }\href
  {https://doi.org/10.1103/PhysRevB.58.R1699} {\bibfield  {journal} {\bibinfo
  {journal} {Phys. Rev. B}\ }\textbf {\bibinfo {volume} {58}},\ \bibinfo
  {pages} {R1699} (\bibinfo {year} {1998})}\BibitemShut {NoStop}%
\bibitem [{\citenamefont {Naiman}\ \emph {et~al.}(1965)\citenamefont {Naiman},
  \citenamefont {Gilmore}, \citenamefont {Dibartolo}, \citenamefont {Linz},\
  and\ \citenamefont {Santoro}}]{Naiman1965}%
  \BibitemOpen
  \bibfield  {author} {\bibinfo {author} {\bibfnamefont {C.~S.}\ \bibnamefont
  {Naiman}}, \bibinfo {author} {\bibfnamefont {R.}~\bibnamefont {Gilmore}},
  \bibinfo {author} {\bibfnamefont {B.}~\bibnamefont {Dibartolo}}, \bibinfo
  {author} {\bibfnamefont {A.}~\bibnamefont {Linz}},\ and\ \bibinfo {author}
  {\bibfnamefont {R.}~\bibnamefont {Santoro}},\ }\href
  {https://doi.org/10.1063/1.1714092} {\bibfield  {journal} {\bibinfo
  {journal} {J. Appl. Phys.}\ }\textbf {\bibinfo {volume} {36}},\ \bibinfo
  {pages} {1044} (\bibinfo {year} {1965})}\BibitemShut {NoStop}%
\bibitem [{\citenamefont {Ren}\ \emph {et~al.}(2011)\citenamefont {Ren},
  \citenamefont {Yan}, \citenamefont {Zhou}, \citenamefont {Goodenough},
  \citenamefont {Jorgensen}, \citenamefont {Short}, \citenamefont {Kim},
  \citenamefont {Proffen}, \citenamefont {Chang},\ and\ \citenamefont
  {McQueeney}}]{Ren2011}%
  \BibitemOpen
  \bibfield  {author} {\bibinfo {author} {\bibfnamefont {Y.}~\bibnamefont
  {Ren}}, \bibinfo {author} {\bibfnamefont {J.-Q.}\ \bibnamefont {Yan}},
  \bibinfo {author} {\bibfnamefont {J.-S.}\ \bibnamefont {Zhou}}, \bibinfo
  {author} {\bibfnamefont {J.~B.}\ \bibnamefont {Goodenough}}, \bibinfo
  {author} {\bibfnamefont {J.~D.}\ \bibnamefont {Jorgensen}}, \bibinfo {author}
  {\bibfnamefont {S.}~\bibnamefont {Short}}, \bibinfo {author} {\bibfnamefont
  {H.}~\bibnamefont {Kim}}, \bibinfo {author} {\bibfnamefont {T.}~\bibnamefont
  {Proffen}}, \bibinfo {author} {\bibfnamefont {S.}~\bibnamefont {Chang}},\
  and\ \bibinfo {author} {\bibfnamefont {R.~J.}\ \bibnamefont {McQueeney}},\
  }\href {https://doi.org/10.1103/PhysRevB.84.214409} {\bibfield  {journal}
  {\bibinfo  {journal} {Phys. Rev. B}\ }\textbf {\bibinfo {volume} {84}},\
  \bibinfo {pages} {214409} (\bibinfo {year} {2011})}\BibitemShut {NoStop}%
\bibitem [{\citenamefont {Hammer}\ \emph {et~al.}(2004)\citenamefont {Hammer},
  \citenamefont {Wu},\ and\ \citenamefont {Leighton}}]{Hammer2004}%
  \BibitemOpen
  \bibfield  {author} {\bibinfo {author} {\bibfnamefont {D.}~\bibnamefont
  {Hammer}}, \bibinfo {author} {\bibfnamefont {J.}~\bibnamefont {Wu}},\ and\
  \bibinfo {author} {\bibfnamefont {C.}~\bibnamefont {Leighton}},\ }\href
  {https://doi.org/10.1103/PhysRevB.69.134407} {\bibfield  {journal} {\bibinfo
  {journal} {Phys. Rev. B}\ }\textbf {\bibinfo {volume} {69}},\ \bibinfo
  {pages} {134407} (\bibinfo {year} {2004})}\BibitemShut {NoStop}%
\bibitem [{\citenamefont {Narasimhan}\ \emph {et~al.}(1985)\citenamefont
  {Narasimhan}, \citenamefont {Keer},\ and\ \citenamefont
  {Chakrabarty}}]{narasimhan1985structural}%
  \BibitemOpen
  \bibfield  {author} {\bibinfo {author} {\bibfnamefont {V.}~\bibnamefont
  {Narasimhan}}, \bibinfo {author} {\bibfnamefont {H.~V.}\ \bibnamefont
  {Keer}},\ and\ \bibinfo {author} {\bibfnamefont {D.~K.}\ \bibnamefont
  {Chakrabarty}},\ }\href
  {https://doi.org/https://doi.org/10.1002/pssa.2210890105} {\bibfield
  {journal} {\bibinfo  {journal} {Phys. Status Solidi A}\ }\textbf {\bibinfo
  {volume} {89}},\ \bibinfo {pages} {65} (\bibinfo {year} {1985})}\BibitemShut
  {NoStop}%
\bibitem [{\citenamefont {Viswanathan}\ and\ \citenamefont
  {Anil~Kumar}(2009)}]{Viswanathan2009}%
  \BibitemOpen
  \bibfield  {author} {\bibinfo {author} {\bibfnamefont {M.}~\bibnamefont
  {Viswanathan}}\ and\ \bibinfo {author} {\bibfnamefont {P.~S.}\ \bibnamefont
  {Anil~Kumar}},\ }\href {https://doi.org/10.1103/PhysRevB.80.012410}
  {\bibfield  {journal} {\bibinfo  {journal} {Phys. Rev. B}\ }\textbf {\bibinfo
  {volume} {80}},\ \bibinfo {pages} {012410(R)} (\bibinfo {year}
  {2009})}\BibitemShut {NoStop}%
\bibitem [{\citenamefont {P{\'{e}}rez}\ \emph {et~al.}(1998)\citenamefont
  {P{\'{e}}rez}, \citenamefont {Garc{\'{i}}a}, \citenamefont {Blasco},\ and\
  \citenamefont {Stankiewicz}}]{Perez1998}%
  \BibitemOpen
  \bibfield  {author} {\bibinfo {author} {\bibfnamefont {J.}~\bibnamefont
  {P{\'{e}}rez}}, \bibinfo {author} {\bibfnamefont {J.}~\bibnamefont
  {Garc{\'{i}}a}}, \bibinfo {author} {\bibfnamefont {J.}~\bibnamefont
  {Blasco}},\ and\ \bibinfo {author} {\bibfnamefont {J.}~\bibnamefont
  {Stankiewicz}},\ }\href {https://doi.org/10.1103/PhysRevLett.80.2401}
  {\bibfield  {journal} {\bibinfo  {journal} {Phys. Rev. Lett.}\ }\textbf
  {\bibinfo {volume} {80}},\ \bibinfo {pages} {2401} (\bibinfo {year}
  {1998})}\BibitemShut {NoStop}%
\bibitem [{\citenamefont {Jonker}\ and\ \citenamefont
  {Van~Santen}(1953)}]{Jonker1953MagneticCobalt}%
  \BibitemOpen
  \bibfield  {author} {\bibinfo {author} {\bibfnamefont {G.~H.}\ \bibnamefont
  {Jonker}}\ and\ \bibinfo {author} {\bibfnamefont {J.~H.}\ \bibnamefont
  {Van~Santen}},\ }\href {https://doi.org/10.1016/S0031-8914(53)80011-X}
  {\bibfield  {journal} {\bibinfo  {journal} {Physica}\ }\textbf {\bibinfo
  {volume} {19}},\ \bibinfo {pages} {120} (\bibinfo {year} {1953})}\BibitemShut
  {NoStop}%
\bibitem [{\citenamefont {Motohashi}\ \emph {et~al.}(2005)\citenamefont
  {Motohashi}, \citenamefont {Caignaert}, \citenamefont {Pralong},
  \citenamefont {Hervieu}, \citenamefont {Maignan},\ and\ \citenamefont
  {Raveau}}]{motohashi2005}%
  \BibitemOpen
  \bibfield  {author} {\bibinfo {author} {\bibfnamefont {T.}~\bibnamefont
  {Motohashi}}, \bibinfo {author} {\bibfnamefont {V.}~\bibnamefont
  {Caignaert}}, \bibinfo {author} {\bibfnamefont {V.}~\bibnamefont {Pralong}},
  \bibinfo {author} {\bibfnamefont {M.}~\bibnamefont {Hervieu}}, \bibinfo
  {author} {\bibfnamefont {A.}~\bibnamefont {Maignan}},\ and\ \bibinfo {author}
  {\bibfnamefont {B.}~\bibnamefont {Raveau}},\ }\href
  {https://doi.org/10.1103/PhysRevB.71.214424} {\bibfield  {journal} {\bibinfo
  {journal} {Phys. Rev. B}\ }\textbf {\bibinfo {volume} {71}},\ \bibinfo
  {pages} {214424} (\bibinfo {year} {2005})}\BibitemShut {NoStop}%
\bibitem [{\citenamefont {Asaba}\ \emph {et~al.}(2018)\citenamefont {Asaba},
  \citenamefont {Xiang}, \citenamefont {Kim}, \citenamefont {Rzchowski},
  \citenamefont {Eom},\ and\ \citenamefont {Li}}]{Asaba2018}%
  \BibitemOpen
  \bibfield  {author} {\bibinfo {author} {\bibfnamefont {T.}~\bibnamefont
  {Asaba}}, \bibinfo {author} {\bibfnamefont {Z.}~\bibnamefont {Xiang}},
  \bibinfo {author} {\bibfnamefont {T.~H.}\ \bibnamefont {Kim}}, \bibinfo
  {author} {\bibfnamefont {M.~S.}\ \bibnamefont {Rzchowski}}, \bibinfo {author}
  {\bibfnamefont {C.~B.}\ \bibnamefont {Eom}},\ and\ \bibinfo {author}
  {\bibfnamefont {L.}~\bibnamefont {Li}},\ }\href
  {https://doi.org/10.1103/PhysRevB.98.121105} {\bibfield  {journal} {\bibinfo
  {journal} {Phys. Rev. B}\ }\textbf {\bibinfo {volume} {98}},\ \bibinfo
  {pages} {121105} (\bibinfo {year} {2018})}\BibitemShut {NoStop}%
\bibitem [{\citenamefont {Millis}\ \emph {et~al.}(1998)\citenamefont {Millis},
  \citenamefont {Darling},\ and\ \citenamefont {Migliori}}]{Millis1998a}%
  \BibitemOpen
  \bibfield  {author} {\bibinfo {author} {\bibfnamefont {A.~J.}\ \bibnamefont
  {Millis}}, \bibinfo {author} {\bibfnamefont {T.}~\bibnamefont {Darling}},\
  and\ \bibinfo {author} {\bibfnamefont {A.}~\bibnamefont {Migliori}},\ }\href
  {https://doi.org/10.1063/1.367310} {\bibfield  {journal} {\bibinfo  {journal}
  {J. Appl. Phys.}\ }\textbf {\bibinfo {volume} {83}},\ \bibinfo {pages} {1588}
  (\bibinfo {year} {1998})}\BibitemShut {NoStop}%
\bibitem [{\citenamefont {Nishio}\ and\ \citenamefont
  {Tsuchiya}(2018)}]{nishio2018sol}%
  \BibitemOpen
  \bibfield  {author} {\bibinfo {author} {\bibfnamefont {K.}~\bibnamefont
  {Nishio}}\ and\ \bibinfo {author} {\bibfnamefont {T.}~\bibnamefont
  {Tsuchiya}},\ }\bibinfo {title} {Sol-gel processing of thin films with metal
  salts},\ in\ \href {https://doi.org/10.1007/978-3-319-32101-1_3} {\emph
  {\bibinfo {booktitle} {Handbook of Sol-Gel Science and Technology}}}\
  (\bibinfo  {publisher} {Springer International Publishing},\ \bibinfo {year}
  {2018})\ pp.\ \bibinfo {pages} {133--154}\BibitemShut {NoStop}%
\bibitem [{\citenamefont {Barla}\ \emph {et~al.}(2016)\citenamefont {Barla},
  \citenamefont {Nicol{\'{a}}s}, \citenamefont {Cocco}, \citenamefont
  {Valvidares}, \citenamefont {Herrero-Mart{\'\i}n}, \citenamefont {Gargiani},
  \citenamefont {Moldes}, \citenamefont {Ruget}, \citenamefont {Pellegrin},\
  and\ \citenamefont {Ferrer}}]{barla2016design}%
  \BibitemOpen
  \bibfield  {author} {\bibinfo {author} {\bibfnamefont {A.}~\bibnamefont
  {Barla}}, \bibinfo {author} {\bibfnamefont {J.}~\bibnamefont
  {Nicol{\'{a}}s}}, \bibinfo {author} {\bibfnamefont {D.}~\bibnamefont
  {Cocco}}, \bibinfo {author} {\bibfnamefont {S.~M.}\ \bibnamefont
  {Valvidares}}, \bibinfo {author} {\bibfnamefont {J.}~\bibnamefont
  {Herrero-Mart{\'\i}n}}, \bibinfo {author} {\bibfnamefont {P.}~\bibnamefont
  {Gargiani}}, \bibinfo {author} {\bibfnamefont {J.}~\bibnamefont {Moldes}},
  \bibinfo {author} {\bibfnamefont {C.}~\bibnamefont {Ruget}}, \bibinfo
  {author} {\bibfnamefont {E.}~\bibnamefont {Pellegrin}},\ and\ \bibinfo
  {author} {\bibfnamefont {S.}~\bibnamefont {Ferrer}},\ }\href
  {https://doi.org/10.1107/S1600577516013461} {\bibfield  {journal} {\bibinfo
  {journal} {J. Synchrotron Radiat.}\ }\textbf {\bibinfo {volume} {23}},\
  \bibinfo {pages} {1507} (\bibinfo {year} {2016})}\BibitemShut {NoStop}%
\bibitem [{\citenamefont {Bj{\"{o}}rck}\ and\ \citenamefont
  {Andersson}(2007)}]{Bjorck2007}%
  \BibitemOpen
  \bibfield  {author} {\bibinfo {author} {\bibfnamefont {M.}~\bibnamefont
  {Bj{\"{o}}rck}}\ and\ \bibinfo {author} {\bibfnamefont {G.}~\bibnamefont
  {Andersson}},\ }\href {https://doi.org/10.1107/S0021889807045086} {\bibfield
  {journal} {\bibinfo  {journal} {J. Appl. Cryst.}\ }\textbf {\bibinfo {volume}
  {40}},\ \bibinfo {pages} {1174} (\bibinfo {year} {2007})}\BibitemShut
  {NoStop}%
\bibitem [{\citenamefont {Wang}\ and\ \citenamefont
  {Lindelof}(1988)}]{Wang1988}%
  \BibitemOpen
  \bibfield  {author} {\bibinfo {author} {\bibfnamefont {S.}~\bibnamefont
  {Wang}}\ and\ \bibinfo {author} {\bibfnamefont {P.~E.}\ \bibnamefont
  {Lindelof}},\ }\href {https://doi.org/10.1007/BF00116871} {\bibfield
  {journal} {\bibinfo  {journal} {J. Low Temp. Phys.}\ }\textbf {\bibinfo
  {volume} {71}},\ \bibinfo {pages} {403} (\bibinfo {year} {1988})}\BibitemShut
  {NoStop}%
\bibitem [{\citenamefont {Hikami}\ \emph {et~al.}(1980)\citenamefont {Hikami},
  \citenamefont {Larkin},\ and\ \citenamefont {Nagaoka}}]{Hikami1980}%
  \BibitemOpen
  \bibfield  {author} {\bibinfo {author} {\bibfnamefont {S.}~\bibnamefont
  {Hikami}}, \bibinfo {author} {\bibfnamefont {A.~I.}\ \bibnamefont {Larkin}},\
  and\ \bibinfo {author} {\bibfnamefont {Y.}~\bibnamefont {Nagaoka}},\ }\href
  {https://doi.org/10.1143/ptp.63.707} {\bibfield  {journal} {\bibinfo
  {journal} {Prog. Theor. Phys.}\ }\textbf {\bibinfo {volume} {63}},\ \bibinfo
  {pages} {707} (\bibinfo {year} {1980})}\BibitemShut {NoStop}%
\bibitem [{\citenamefont {Wei}\ \emph {et~al.}(1988)\citenamefont {Wei},
  \citenamefont {Bergmann},\ and\ \citenamefont {Peters}}]{Wei1988}%
  \BibitemOpen
  \bibfield  {author} {\bibinfo {author} {\bibfnamefont {W.}~\bibnamefont
  {Wei}}, \bibinfo {author} {\bibfnamefont {G.}~\bibnamefont {Bergmann}},\ and\
  \bibinfo {author} {\bibfnamefont {R.~P.}\ \bibnamefont {Peters}},\ }\href
  {https://doi.org/10.1103/PhysRevB.38.11751} {\bibfield  {journal} {\bibinfo
  {journal} {Phys. Rev. B}\ }\textbf {\bibinfo {volume} {38}},\ \bibinfo
  {pages} {11751} (\bibinfo {year} {1988})}\BibitemShut {NoStop}%
\bibitem [{\citenamefont {Peters}\ \emph {et~al.}(1988)\citenamefont {Peters},
  \citenamefont {Bergmann},\ and\ \citenamefont {Mueller}}]{Peters1988}%
  \BibitemOpen
  \bibfield  {author} {\bibinfo {author} {\bibfnamefont {R.~P.}\ \bibnamefont
  {Peters}}, \bibinfo {author} {\bibfnamefont {G.}~\bibnamefont {Bergmann}},\
  and\ \bibinfo {author} {\bibfnamefont {R.~M.}\ \bibnamefont {Mueller}},\
  }\href {https://doi.org/10.1103/PhysRevLett.60.1093} {\bibfield  {journal}
  {\bibinfo  {journal} {Phys. Rev. Lett.}\ }\textbf {\bibinfo {volume} {60}},\
  \bibinfo {pages} {1093} (\bibinfo {year} {1988})}\BibitemShut {NoStop}%
\bibitem [{\citenamefont {Schopfer}\ \emph {et~al.}(2003)\citenamefont
  {Schopfer}, \citenamefont {B\"auerle}, \citenamefont {Rabaud},\ and\
  \citenamefont {Saminadayar}}]{Schopfer2003}%
  \BibitemOpen
  \bibfield  {author} {\bibinfo {author} {\bibfnamefont {F.}~\bibnamefont
  {Schopfer}}, \bibinfo {author} {\bibfnamefont {C.}~\bibnamefont {B\"auerle}},
  \bibinfo {author} {\bibfnamefont {W.}~\bibnamefont {Rabaud}},\ and\ \bibinfo
  {author} {\bibfnamefont {L.}~\bibnamefont {Saminadayar}},\ }\href
  {https://doi.org/10.1103/PhysRevLett.90.056801} {\bibfield  {journal}
  {\bibinfo  {journal} {Phys. Rev. Lett.}\ }\textbf {\bibinfo {volume} {90}},\
  \bibinfo {pages} {056801} (\bibinfo {year} {2003})}\BibitemShut {NoStop}%
\bibitem [{\citenamefont {Rushforth}\ \emph {et~al.}(2007)\citenamefont
  {Rushforth}, \citenamefont {V\'yborn\'y}, \citenamefont {King}, \citenamefont
  {Edmonds}, \citenamefont {Campion}, \citenamefont {Foxon}, \citenamefont
  {Wunderlich}, \citenamefont {Irvine}, \citenamefont
  {Va\ifmmode~\check{s}\else \v{s}\fi{}ek}, \citenamefont {Nov\'ak},
  \citenamefont {Olejn\'{\i}k}, \citenamefont {Sinova}, \citenamefont
  {Jungwirth},\ and\ \citenamefont {Gallagher}}]{Rushforth2007}%
  \BibitemOpen
  \bibfield  {author} {\bibinfo {author} {\bibfnamefont {A.~W.}\ \bibnamefont
  {Rushforth}}, \bibinfo {author} {\bibfnamefont {K.}~\bibnamefont
  {V\'yborn\'y}}, \bibinfo {author} {\bibfnamefont {C.~S.}\ \bibnamefont
  {King}}, \bibinfo {author} {\bibfnamefont {K.~W.}\ \bibnamefont {Edmonds}},
  \bibinfo {author} {\bibfnamefont {R.~P.}\ \bibnamefont {Campion}}, \bibinfo
  {author} {\bibfnamefont {C.~T.}\ \bibnamefont {Foxon}}, \bibinfo {author}
  {\bibfnamefont {J.}~\bibnamefont {Wunderlich}}, \bibinfo {author}
  {\bibfnamefont {A.~C.}\ \bibnamefont {Irvine}}, \bibinfo {author}
  {\bibfnamefont {P.}~\bibnamefont {Va\ifmmode~\check{s}\else \v{s}\fi{}ek}},
  \bibinfo {author} {\bibfnamefont {V.}~\bibnamefont {Nov\'ak}}, \bibinfo
  {author} {\bibfnamefont {K.}~\bibnamefont {Olejn\'{\i}k}}, \bibinfo {author}
  {\bibfnamefont {J.}~\bibnamefont {Sinova}}, \bibinfo {author} {\bibfnamefont
  {T.}~\bibnamefont {Jungwirth}},\ and\ \bibinfo {author} {\bibfnamefont
  {B.~L.}\ \bibnamefont {Gallagher}},\ }\href
  {https://doi.org/10.1103/PhysRevLett.99.147207} {\bibfield  {journal}
  {\bibinfo  {journal} {Phys. Rev. Lett.}\ }\textbf {\bibinfo {volume} {99}},\
  \bibinfo {pages} {147207} (\bibinfo {year} {2007})}\BibitemShut {NoStop}%
\bibitem [{\citenamefont {Kartsovnik}\ \emph {et~al.}(1992)\citenamefont
  {Kartsovnik}, \citenamefont {Laukhin}, \citenamefont {Pesotskii},
  \citenamefont {Schegolev},\ and\ \citenamefont {Yakovenko}}]{Kartsovnik1992}%
  \BibitemOpen
  \bibfield  {author} {\bibinfo {author} {\bibfnamefont {M.~V.}\ \bibnamefont
  {Kartsovnik}}, \bibinfo {author} {\bibfnamefont {V.~N.}\ \bibnamefont
  {Laukhin}}, \bibinfo {author} {\bibfnamefont {S.~I.}\ \bibnamefont
  {Pesotskii}}, \bibinfo {author} {\bibfnamefont {I.~F.}\ \bibnamefont
  {Schegolev}},\ and\ \bibinfo {author} {\bibfnamefont {V.~M.}\ \bibnamefont
  {Yakovenko}},\ }\href {https://doi.org/10.1051/jp1:1992125} {\bibfield
  {journal} {\bibinfo  {journal} {J. phys., I}\ }\textbf {\bibinfo {volume}
  {2}},\ \bibinfo {pages} {89} (\bibinfo {year} {1992})}\BibitemShut {NoStop}%
\bibitem [{\citenamefont {Wu}\ \emph {et~al.}(2008)\citenamefont {Wu},
  \citenamefont {Wei}, \citenamefont {Johnston-Halperin}, \citenamefont
  {Awschalom},\ and\ \citenamefont {Shi}}]{Wu1253}%
  \BibitemOpen
  \bibfield  {author} {\bibinfo {author} {\bibfnamefont {D.}~\bibnamefont
  {Wu}}, \bibinfo {author} {\bibfnamefont {P.}~\bibnamefont {Wei}}, \bibinfo
  {author} {\bibfnamefont {E.}~\bibnamefont {Johnston-Halperin}}, \bibinfo
  {author} {\bibfnamefont {D.~D.}\ \bibnamefont {Awschalom}},\ and\ \bibinfo
  {author} {\bibfnamefont {J.}~\bibnamefont {Shi}},\ }\href
  {https://doi.org/10.1103/PhysRevB.77.125320} {\bibfield  {journal} {\bibinfo
  {journal} {Phys. Rev. B}\ }\textbf {\bibinfo {volume} {77}},\ \bibinfo
  {pages} {125320} (\bibinfo {year} {2008})}\BibitemShut {NoStop}%
\bibitem [{\citenamefont {Li}\ \emph {et~al.}(2010)\citenamefont {Li},
  \citenamefont {Jin}, \citenamefont {Jiang},\ and\ \citenamefont
  {Bai}}]{Li2010a}%
  \BibitemOpen
  \bibfield  {author} {\bibinfo {author} {\bibfnamefont {P.}~\bibnamefont
  {Li}}, \bibinfo {author} {\bibfnamefont {C.}~\bibnamefont {Jin}}, \bibinfo
  {author} {\bibfnamefont {E.~Y.}\ \bibnamefont {Jiang}},\ and\ \bibinfo
  {author} {\bibfnamefont {H.~L.}\ \bibnamefont {Bai}},\ }\href
  {https://doi.org/10.1063/1.3499696} {\bibfield  {journal} {\bibinfo
  {journal} {J. Appl. Phys.}\ }\textbf {\bibinfo {volume} {108}},\ \bibinfo
  {pages} {093921} (\bibinfo {year} {2010})}\BibitemShut {NoStop}%
\bibitem [{\citenamefont {Chen}\ \emph {et~al.}(2009)\citenamefont {Chen},
  \citenamefont {Sun}, \citenamefont {Zhao}, \citenamefont {Wang},
  \citenamefont {Wang}, \citenamefont {Shen},\ and\ \citenamefont
  {Pryds}}]{Chen2009}%
  \BibitemOpen
  \bibfield  {author} {\bibinfo {author} {\bibfnamefont {Y.~Z.}\ \bibnamefont
  {Chen}}, \bibinfo {author} {\bibfnamefont {J.~R.}\ \bibnamefont {Sun}},
  \bibinfo {author} {\bibfnamefont {T.~Y.}\ \bibnamefont {Zhao}}, \bibinfo
  {author} {\bibfnamefont {J.}~\bibnamefont {Wang}}, \bibinfo {author}
  {\bibfnamefont {Z.~H.}\ \bibnamefont {Wang}}, \bibinfo {author}
  {\bibfnamefont {B.~G.}\ \bibnamefont {Shen}},\ and\ \bibinfo {author}
  {\bibfnamefont {N.}~\bibnamefont {Pryds}},\ }\href
  {https://doi.org/10.1063/1.3240407} {\bibfield  {journal} {\bibinfo
  {journal} {Appl. Phys. Lett.}\ }\textbf {\bibinfo {volume} {95}},\ \bibinfo
  {pages} {132506} (\bibinfo {year} {2009})}\BibitemShut {NoStop}%
\bibitem [{\citenamefont {Malozemoff}(1985)}]{Malozemoff1985}%
  \BibitemOpen
  \bibfield  {author} {\bibinfo {author} {\bibfnamefont {A.~P.}\ \bibnamefont
  {Malozemoff}},\ }\href {https://doi.org/10.1103/PhysRevB.32.6080} {\bibfield
  {journal} {\bibinfo  {journal} {Phys. Rev. B}\ }\textbf {\bibinfo {volume}
  {32}},\ \bibinfo {pages} {6080} (\bibinfo {year} {1985})}\BibitemShut
  {NoStop}%
\bibitem [{\citenamefont {Campbell}\ \emph {et~al.}(1970)\citenamefont
  {Campbell}, \citenamefont {Fert},\ and\ \citenamefont
  {Jaoul}}]{Campbell1970}%
  \BibitemOpen
  \bibfield  {author} {\bibinfo {author} {\bibfnamefont {I.~A.}\ \bibnamefont
  {Campbell}}, \bibinfo {author} {\bibfnamefont {A.}~\bibnamefont {Fert}},\
  and\ \bibinfo {author} {\bibfnamefont {O.}~\bibnamefont {Jaoul}},\ }\href
  {https://doi.org/10.1088/0022-3719/3/1S/310} {\bibfield  {journal} {\bibinfo
  {journal} {J. Phys. C: Solid State Phys.}\ }\textbf {\bibinfo {volume} {3}},\
  \bibinfo {pages} {S95} (\bibinfo {year} {1970})}\BibitemShut {NoStop}%
\bibitem [{\citenamefont {Kokado}\ \emph {et~al.}(2012)\citenamefont {Kokado},
  \citenamefont {Tsunoda}, \citenamefont {Harigaya},\ and\ \citenamefont
  {Sakuma}}]{Kokado2012a}%
  \BibitemOpen
  \bibfield  {author} {\bibinfo {author} {\bibfnamefont {S.}~\bibnamefont
  {Kokado}}, \bibinfo {author} {\bibfnamefont {M.}~\bibnamefont {Tsunoda}},
  \bibinfo {author} {\bibfnamefont {K.}~\bibnamefont {Harigaya}},\ and\
  \bibinfo {author} {\bibfnamefont {A.}~\bibnamefont {Sakuma}},\ }\href
  {https://doi.org/10.1143/JPSJ.81.024705} {\bibfield  {journal} {\bibinfo
  {journal} {J. Phys. Soc. Jpn.}\ }\textbf {\bibinfo {volume} {81}},\ \bibinfo
  {pages} {024705} (\bibinfo {year} {2012})}\BibitemShut {NoStop}%
\bibitem [{\citenamefont {Ky\^omen}\ \emph
  {et~al.}(2003{\natexlab{a}})\citenamefont {Ky\^omen}, \citenamefont
  {Yamazaki},\ and\ \citenamefont {Itoh}}]{Kyomen2003}%
  \BibitemOpen
  \bibfield  {author} {\bibinfo {author} {\bibfnamefont {T.}~\bibnamefont
  {Ky\^omen}}, \bibinfo {author} {\bibfnamefont {R.}~\bibnamefont {Yamazaki}},\
  and\ \bibinfo {author} {\bibfnamefont {M.}~\bibnamefont {Itoh}},\ }\href
  {https://doi.org/10.1103/PhysRevB.68.104416} {\bibfield  {journal} {\bibinfo
  {journal} {Phys. Rev. B}\ }\textbf {\bibinfo {volume} {68}},\ \bibinfo
  {pages} {104416} (\bibinfo {year} {2003}{\natexlab{a}})}\BibitemShut
  {NoStop}%
\bibitem [{\citenamefont {Ky\^omen}\ \emph
  {et~al.}(2003{\natexlab{b}})\citenamefont {Ky\^omen}, \citenamefont {Asaka},\
  and\ \citenamefont {Itoh}}]{Kyomen2003b}%
  \BibitemOpen
  \bibfield  {author} {\bibinfo {author} {\bibfnamefont {T.}~\bibnamefont
  {Ky\^omen}}, \bibinfo {author} {\bibfnamefont {Y.}~\bibnamefont {Asaka}},\
  and\ \bibinfo {author} {\bibfnamefont {M.}~\bibnamefont {Itoh}},\ }\href
  {https://doi.org/10.1103/PhysRevB.67.144424} {\bibfield  {journal} {\bibinfo
  {journal} {Phys. Rev. B}\ }\textbf {\bibinfo {volume} {67}},\ \bibinfo
  {pages} {144424} (\bibinfo {year} {2003}{\natexlab{b}})}\BibitemShut
  {NoStop}%
\bibitem [{\citenamefont {Lin}\ \emph {et~al.}(2010)\citenamefont {Lin},
  \citenamefont {Chin}, \citenamefont {Hu}, \citenamefont {Shu}, \citenamefont
  {Chou}, \citenamefont {Ohta}, \citenamefont {Yoshimura}, \citenamefont
  {H\'ebert}, \citenamefont {Maignan}, \citenamefont {Tanaka}, \citenamefont
  {Tjeng},\ and\ \citenamefont {Chen}}]{lin2010local}%
  \BibitemOpen
  \bibfield  {author} {\bibinfo {author} {\bibfnamefont {H.-J.}\ \bibnamefont
  {Lin}}, \bibinfo {author} {\bibfnamefont {Y.~Y.}\ \bibnamefont {Chin}},
  \bibinfo {author} {\bibfnamefont {Z.}~\bibnamefont {Hu}}, \bibinfo {author}
  {\bibfnamefont {G.~J.}\ \bibnamefont {Shu}}, \bibinfo {author} {\bibfnamefont
  {F.~C.}\ \bibnamefont {Chou}}, \bibinfo {author} {\bibfnamefont
  {H.}~\bibnamefont {Ohta}}, \bibinfo {author} {\bibfnamefont {K.}~\bibnamefont
  {Yoshimura}}, \bibinfo {author} {\bibfnamefont {S.}~\bibnamefont {H\'ebert}},
  \bibinfo {author} {\bibfnamefont {A.}~\bibnamefont {Maignan}}, \bibinfo
  {author} {\bibfnamefont {A.}~\bibnamefont {Tanaka}}, \bibinfo {author}
  {\bibfnamefont {L.~H.}\ \bibnamefont {Tjeng}},\ and\ \bibinfo {author}
  {\bibfnamefont {C.~T.}\ \bibnamefont {Chen}},\ }\href
  {https://doi.org/10.1103/PhysRevB.81.115138} {\bibfield  {journal} {\bibinfo
  {journal} {Phys. Rev. B}\ }\textbf {\bibinfo {volume} {81}},\ \bibinfo
  {pages} {115138} (\bibinfo {year} {2010})}\BibitemShut {NoStop}%
\bibitem [{\citenamefont {Guillou}\ \emph {et~al.}(2017)\citenamefont
  {Guillou}, \citenamefont {Kummer}, \citenamefont {Br\'eard}, \citenamefont
  {Herv\'e},\ and\ \citenamefont {Hardy}}]{guillou2017valence}%
  \BibitemOpen
  \bibfield  {author} {\bibinfo {author} {\bibfnamefont {F.}~\bibnamefont
  {Guillou}}, \bibinfo {author} {\bibfnamefont {K.}~\bibnamefont {Kummer}},
  \bibinfo {author} {\bibfnamefont {Y.}~\bibnamefont {Br\'eard}}, \bibinfo
  {author} {\bibfnamefont {L.}~\bibnamefont {Herv\'e}},\ and\ \bibinfo {author}
  {\bibfnamefont {V.}~\bibnamefont {Hardy}},\ }\href
  {https://doi.org/10.1103/PhysRevB.95.174445} {\bibfield  {journal} {\bibinfo
  {journal} {Phys. Rev. B}\ }\textbf {\bibinfo {volume} {95}},\ \bibinfo
  {pages} {174445} (\bibinfo {year} {2017})}\BibitemShut {NoStop}%
\bibitem [{\citenamefont {Liao}\ \emph {et~al.}(2019)\citenamefont {Liao},
  \citenamefont {Haw}, \citenamefont {Kuo}, \citenamefont {Guo}, \citenamefont
  {Vasili}, \citenamefont {Valvidares}, \citenamefont {Komarek}, \citenamefont
  {Ishii}, \citenamefont {Chen}, \citenamefont {Lin}, \citenamefont {Tanaka},
  \citenamefont {Chan}, \citenamefont {Tjeng}, \citenamefont {Chen},
  \citenamefont {Hu},\ and\ \citenamefont {Chen}}]{liao2019valence}%
  \BibitemOpen
  \bibfield  {author} {\bibinfo {author} {\bibfnamefont {S.-C.}\ \bibnamefont
  {Liao}}, \bibinfo {author} {\bibfnamefont {S.~C.}\ \bibnamefont {Haw}},
  \bibinfo {author} {\bibfnamefont {C.~Y.}\ \bibnamefont {Kuo}}, \bibinfo
  {author} {\bibfnamefont {H.}~\bibnamefont {Guo}}, \bibinfo {author}
  {\bibfnamefont {H.~B.}\ \bibnamefont {Vasili}}, \bibinfo {author}
  {\bibfnamefont {S.~M.}\ \bibnamefont {Valvidares}}, \bibinfo {author}
  {\bibfnamefont {A.~C.}\ \bibnamefont {Komarek}}, \bibinfo {author}
  {\bibfnamefont {H.}~\bibnamefont {Ishii}}, \bibinfo {author} {\bibfnamefont
  {S.~A.}\ \bibnamefont {Chen}}, \bibinfo {author} {\bibfnamefont {H.~J.}\
  \bibnamefont {Lin}}, \bibinfo {author} {\bibfnamefont {A.}~\bibnamefont
  {Tanaka}}, \bibinfo {author} {\bibfnamefont {T.~S.}\ \bibnamefont {Chan}},
  \bibinfo {author} {\bibfnamefont {L.~H.}\ \bibnamefont {Tjeng}}, \bibinfo
  {author} {\bibfnamefont {C.~T.}\ \bibnamefont {Chen}}, \bibinfo {author}
  {\bibfnamefont {Z.}~\bibnamefont {Hu}},\ and\ \bibinfo {author}
  {\bibfnamefont {J.~M.}\ \bibnamefont {Chen}},\ }\href
  {https://doi.org/10.1103/PhysRevB.99.075110} {\bibfield  {journal} {\bibinfo
  {journal} {Phys. Rev. B}\ }\textbf {\bibinfo {volume} {99}},\ \bibinfo
  {pages} {075110} (\bibinfo {year} {2019})}\BibitemShut {NoStop}%
\bibitem [{\citenamefont {Chen}\ \emph {et~al.}(2014)\citenamefont {Chen},
  \citenamefont {Chin}, \citenamefont {Valldor}, \citenamefont {Hu},
  \citenamefont {Lee}, \citenamefont {Haw}, \citenamefont {Hiraoka},
  \citenamefont {Ishii}, \citenamefont {Pao}, \citenamefont {Tsuei},
  \citenamefont {Lee}, \citenamefont {Lin}, \citenamefont {Jang}, \citenamefont
  {Tanaka}, \citenamefont {Chen},\ and\ \citenamefont
  {Tjeng}}]{chen2014complete}%
  \BibitemOpen
  \bibfield  {author} {\bibinfo {author} {\bibfnamefont {J.-M.}\ \bibnamefont
  {Chen}}, \bibinfo {author} {\bibfnamefont {Y.-Y.}\ \bibnamefont {Chin}},
  \bibinfo {author} {\bibfnamefont {M.}~\bibnamefont {Valldor}}, \bibinfo
  {author} {\bibfnamefont {Z.}~\bibnamefont {Hu}}, \bibinfo {author}
  {\bibfnamefont {J.-M.}\ \bibnamefont {Lee}}, \bibinfo {author} {\bibfnamefont
  {S.-C.}\ \bibnamefont {Haw}}, \bibinfo {author} {\bibfnamefont
  {N.}~\bibnamefont {Hiraoka}}, \bibinfo {author} {\bibfnamefont
  {H.}~\bibnamefont {Ishii}}, \bibinfo {author} {\bibfnamefont {C.-W.}\
  \bibnamefont {Pao}}, \bibinfo {author} {\bibfnamefont {K.-D.}\ \bibnamefont
  {Tsuei}}, \bibinfo {author} {\bibfnamefont {J.-F.}\ \bibnamefont {Lee}},
  \bibinfo {author} {\bibfnamefont {H.-J.}\ \bibnamefont {Lin}}, \bibinfo
  {author} {\bibfnamefont {L.-Y.}\ \bibnamefont {Jang}}, \bibinfo {author}
  {\bibfnamefont {A.}~\bibnamefont {Tanaka}}, \bibinfo {author} {\bibfnamefont
  {C.-T.}\ \bibnamefont {Chen}},\ and\ \bibinfo {author} {\bibfnamefont
  {L.~H.}\ \bibnamefont {Tjeng}},\ }\href {https://doi.org/10.1021/ja4114006}
  {\bibfield  {journal} {\bibinfo  {journal} {J. Am. Chem. Soc.}\ }\textbf
  {\bibinfo {volume} {136}},\ \bibinfo {pages} {1514} (\bibinfo {year}
  {2014})}\BibitemShut {NoStop}%
\bibitem [{\citenamefont {Haverkort}\ \emph {et~al.}(2006)\citenamefont
  {Haverkort}, \citenamefont {Hu}, \citenamefont {Cezar}, \citenamefont
  {Burnus}, \citenamefont {Hartmann}, \citenamefont {Reuther}, \citenamefont
  {Zobel}, \citenamefont {Lorenz}, \citenamefont {Tanaka}, \citenamefont
  {Brookes}, \citenamefont {Hsieh}, \citenamefont {Lin}, \citenamefont {Chen},\
  and\ \citenamefont {Tjeng}}]{haverkort2006spin}%
  \BibitemOpen
  \bibfield  {author} {\bibinfo {author} {\bibfnamefont {M.~W.}\ \bibnamefont
  {Haverkort}}, \bibinfo {author} {\bibfnamefont {Z.}~\bibnamefont {Hu}},
  \bibinfo {author} {\bibfnamefont {J.~C.}\ \bibnamefont {Cezar}}, \bibinfo
  {author} {\bibfnamefont {T.}~\bibnamefont {Burnus}}, \bibinfo {author}
  {\bibfnamefont {H.}~\bibnamefont {Hartmann}}, \bibinfo {author}
  {\bibfnamefont {M.}~\bibnamefont {Reuther}}, \bibinfo {author} {\bibfnamefont
  {C.}~\bibnamefont {Zobel}}, \bibinfo {author} {\bibfnamefont
  {T.}~\bibnamefont {Lorenz}}, \bibinfo {author} {\bibfnamefont
  {A.}~\bibnamefont {Tanaka}}, \bibinfo {author} {\bibfnamefont {N.~B.}\
  \bibnamefont {Brookes}}, \bibinfo {author} {\bibfnamefont {H.~H.}\
  \bibnamefont {Hsieh}}, \bibinfo {author} {\bibfnamefont {H.-J.}\ \bibnamefont
  {Lin}}, \bibinfo {author} {\bibfnamefont {C.~T.}\ \bibnamefont {Chen}},\ and\
  \bibinfo {author} {\bibfnamefont {L.~H.}\ \bibnamefont {Tjeng}},\ }\href
  {https://doi.org/10.1103/PhysRevLett.97.176405} {\bibfield  {journal}
  {\bibinfo  {journal} {Phys. Rev. Lett.}\ }\textbf {\bibinfo {volume} {97}},\
  \bibinfo {pages} {176405} (\bibinfo {year} {2006})}\BibitemShut {NoStop}%
\bibitem [{\citenamefont {Shen}\ \emph {et~al.}(1995)\citenamefont {Shen},
  \citenamefont {Luo},\ and\ \citenamefont {Furdyna}}]{shen1995spatial}%
  \BibitemOpen
  \bibfield  {author} {\bibinfo {author} {\bibfnamefont {Q.}~\bibnamefont
  {Shen}}, \bibinfo {author} {\bibfnamefont {H.}~\bibnamefont {Luo}},\ and\
  \bibinfo {author} {\bibfnamefont {J.~K.}\ \bibnamefont {Furdyna}},\ }\href
  {https://doi.org/10.1103/PhysRevLett.75.2590} {\bibfield  {journal} {\bibinfo
   {journal} {Phys. Rev. Lett.}\ }\textbf {\bibinfo {volume} {75}},\ \bibinfo
  {pages} {2590} (\bibinfo {year} {1995})}\BibitemShut {NoStop}%
\bibitem [{\citenamefont {Piamonteze}\ \emph {et~al.}(2009)\citenamefont
  {Piamonteze}, \citenamefont {Miedema},\ and\ \citenamefont
  {de~Groot}}]{piamonteze2009accuracy}%
  \BibitemOpen
  \bibfield  {author} {\bibinfo {author} {\bibfnamefont {C.}~\bibnamefont
  {Piamonteze}}, \bibinfo {author} {\bibfnamefont {P.}~\bibnamefont
  {Miedema}},\ and\ \bibinfo {author} {\bibfnamefont {F.~M.~F.}\ \bibnamefont
  {de~Groot}},\ }\href {https://doi.org/10.1103/PhysRevB.80.184410} {\bibfield
  {journal} {\bibinfo  {journal} {Phys. Rev. B}\ }\textbf {\bibinfo {volume}
  {80}},\ \bibinfo {pages} {184410} (\bibinfo {year} {2009})}\BibitemShut
  {NoStop}%
\bibitem [{\citenamefont {Merz}\ \emph {et~al.}(2010)\citenamefont {Merz},
  \citenamefont {Nagel}, \citenamefont {Pinta}, \citenamefont {Samartsev},
  \citenamefont {v.~L\"ohneysen}, \citenamefont {Wissinger}, \citenamefont
  {Uebe}, \citenamefont {Assmann}, \citenamefont {Fuchs},\ and\ \citenamefont
  {Schuppler}}]{merz2010x}%
  \BibitemOpen
  \bibfield  {author} {\bibinfo {author} {\bibfnamefont {M.}~\bibnamefont
  {Merz}}, \bibinfo {author} {\bibfnamefont {P.}~\bibnamefont {Nagel}},
  \bibinfo {author} {\bibfnamefont {C.}~\bibnamefont {Pinta}}, \bibinfo
  {author} {\bibfnamefont {A.}~\bibnamefont {Samartsev}}, \bibinfo {author}
  {\bibfnamefont {H.}~\bibnamefont {v.~L\"ohneysen}}, \bibinfo {author}
  {\bibfnamefont {M.}~\bibnamefont {Wissinger}}, \bibinfo {author}
  {\bibfnamefont {S.}~\bibnamefont {Uebe}}, \bibinfo {author} {\bibfnamefont
  {A.}~\bibnamefont {Assmann}}, \bibinfo {author} {\bibfnamefont
  {D.}~\bibnamefont {Fuchs}},\ and\ \bibinfo {author} {\bibfnamefont
  {S.}~\bibnamefont {Schuppler}},\ }\href
  {https://doi.org/10.1103/PhysRevB.82.174416} {\bibfield  {journal} {\bibinfo
  {journal} {Phys. Rev. B}\ }\textbf {\bibinfo {volume} {82}},\ \bibinfo
  {pages} {174416} (\bibinfo {year} {2010})}\BibitemShut {NoStop}%
\bibitem [{\citenamefont {Piamonteze}\ \emph {et~al.}(2015)\citenamefont
  {Piamonteze}, \citenamefont {Gibert}, \citenamefont {Heidler}, \citenamefont
  {Dreiser}, \citenamefont {Rusponi}, \citenamefont {Brune}, \citenamefont
  {Triscone}, \citenamefont {Nolting},\ and\ \citenamefont
  {Staub}}]{piamonteze2015interfacial}%
  \BibitemOpen
  \bibfield  {author} {\bibinfo {author} {\bibfnamefont {C.}~\bibnamefont
  {Piamonteze}}, \bibinfo {author} {\bibfnamefont {M.}~\bibnamefont {Gibert}},
  \bibinfo {author} {\bibfnamefont {J.}~\bibnamefont {Heidler}}, \bibinfo
  {author} {\bibfnamefont {J.}~\bibnamefont {Dreiser}}, \bibinfo {author}
  {\bibfnamefont {S.}~\bibnamefont {Rusponi}}, \bibinfo {author} {\bibfnamefont
  {H.}~\bibnamefont {Brune}}, \bibinfo {author} {\bibfnamefont {J.-M.}\
  \bibnamefont {Triscone}}, \bibinfo {author} {\bibfnamefont {F.}~\bibnamefont
  {Nolting}},\ and\ \bibinfo {author} {\bibfnamefont {U.}~\bibnamefont
  {Staub}},\ }\href {https://doi.org/10.1103/PhysRevB.92.014426} {\bibfield
  {journal} {\bibinfo  {journal} {Phys. Rev. B}\ }\textbf {\bibinfo {volume}
  {92}},\ \bibinfo {pages} {014426} (\bibinfo {year} {2015})}\BibitemShut
  {NoStop}%
\bibitem [{\citenamefont {Freeland}\ \emph {et~al.}(2008)\citenamefont
  {Freeland}, \citenamefont {Ma},\ and\ \citenamefont {Shi}}]{freeland2008}%
  \BibitemOpen
  \bibfield  {author} {\bibinfo {author} {\bibfnamefont {J.}~\bibnamefont
  {Freeland}}, \bibinfo {author} {\bibfnamefont {J.}~\bibnamefont {Ma}},\ and\
  \bibinfo {author} {\bibfnamefont {J.}~\bibnamefont {Shi}},\ }\href
  {https://doi.org/10.1063/1.3027063} {\bibfield  {journal} {\bibinfo
  {journal} {Appl. Phys. Lett.}\ }\textbf {\bibinfo {volume} {93}},\ \bibinfo
  {pages} {212501} (\bibinfo {year} {2008})}\BibitemShut {NoStop}%
\bibitem [{\citenamefont {Raychaudhuri}(1995)}]{Raychaudhuri1995}%
  \BibitemOpen
  \bibfield  {author} {\bibinfo {author} {\bibfnamefont {A.~K.}\ \bibnamefont
  {Raychaudhuri}},\ }\href {https://doi.org/10.1080/00018739500101486}
  {\bibfield  {journal} {\bibinfo  {journal} {Adv. Phys.}\ }\textbf {\bibinfo
  {volume} {44}},\ \bibinfo {pages} {21} (\bibinfo {year} {1995})}\BibitemShut
  {NoStop}%
\end{thebibliography}%
